\documentclass{aa}

\usepackage{graphicx}
\usepackage{txfonts}
\usepackage[colorlinks=true,linkcolor=red,citecolor=blue]{hyperref}
\begin{document}

\title{Constraining the origin and models of chemical enrichment in galaxy clusters using the \textsl{Athena} X-IFU}

\author{F.~Mernier\inst{\ref{inst17},\ref{inst16},\ref{inst20},\ref{inst18}} \and E.~Cucchetti\inst{\ref{inst1}} \and L.~Tornatore\inst{\ref{inst3}} \and V.~Biffi\inst{\ref{inst4}, \ref{inst21}} \and E.~Pointecouteau\inst{\ref{inst1}}  \and N.~Clerc\inst{\ref{inst1}} \and P.~Peille\inst{\ref{inst2}}  \and E.~Rasia\inst{\ref{inst3},\ref{inst21}}  \and D.~Barret\inst{\ref{inst1}} \and S.~Borgani\inst{\ref{inst3},\ref{inst5},\ref{inst6},\ref{inst21}}  \and E.~Bulbul\inst{\ref{inst13},\ref{inst14}}  \and T.~Dauser\inst{\ref{inst15}}  \and K.~Dolag\inst{\ref{inst7},\ref{inst8}}  \and S.~Ettori\inst{\ref{inst10},\ref{inst11}} \and M. Gaspari\inst{\ref{inst12},\ref{inst10}}\thanks{\textit{Lyman Spitzer Jr.} Fellow} \and F.~Pajot\inst{\ref{inst1}} \and M.~Roncarelli\inst{\ref{inst10},\ref{inst9}} \and J.~Wilms\inst{\ref{inst15}} 
\and C. No\^us\inst{\ref{inst19}}
}

\institute{ 
European Space Agency (ESA), European Space Research and Technology Centre (ESTEC), Keplerlaan 1, 2201 AZ Noordwijk, The Netherlands \label{inst17} \\ \email{francois.mernier@esa.int}
\and MTA-E\"otv\"os University Lend\"ulet Hot Universe Research Group, P\'azm\'any P\'eter s\'et\'any 1/A, Budapest, 1117, Hungary \label{inst16}
\and Institute of Physics, E\"otv\"os University, P\'azm\'any P\'eter s\'et\'any 1/A, Budapest, 1117, Hungary \label{inst20}
\and SRON Netherlands Institute for Space Research, Sorbonnelaan 2, 3584 CA Utrecht, The Netherlands \label{inst18}
\and IRAP, Université de Toulouse, CNRS, CNES, UPS, Toulouse, France
\label{inst1}
\and INAF, Osservatorio Astronomico di Trieste, via Tiepolo 11, I-34131, Trieste, Italy \label{inst3}
\and Universit\"ats-Sternwarte M\"unchen, Fakult\"at f\"ur Physik, LMU Munich, Scheinerstr. 1, 81679 M\"unchen, Germany \label{inst4}
\and IFPU - Institute for Fundamental Physics of the Universe, Via Beirut 2, 34014 Trieste, Italy \label{inst21}
\and CNES, 18 Avenue Edouard Belin 31400 Toulouse France \label{inst2}
\and Dipartimento di Fisica dell'Università di Trieste, Sezione di Astronomia, via Tiepolo 11, I-34131, Trieste, Italy \label{inst5}
\and INFN - National Institute for Nuclear Physics, Via Valerio 2, I-34127, Trieste Italy \label{inst6}
\and Max-Planck-Institut f\"ur extraterrestrische Physik, Gie{\ss}enbachstra{\ss}e 1, 85748 Garching, Germany  \label{inst13}
\and Harvard-Smithsonian Center for Astrophysics, 60 Garden Street, Cambridge, MA, 02138, USA  \label{inst14}
\and Dr. Karl Remeis-Observatory, Erlangen Centre for Astroparticle Physics, Sternwartstr. 7 96049 Bamberg Germany  \label{inst15}
\and University Observatory Munich, Scheinerstr. 1, D-81679, Munich, Germany \label{inst7}
\and Max Plank Institut für Astrophysik, Karl-Schwarzschield Strasse 1, 85748 Garching bei Munchen, Germany \label{inst8}
\and INAF, Osservatorio di Astrofisica e Scienza dello Spazio, via Piero Gobetti 93/3, 40129 Bologna, Italy \label{inst10}
\and INFN, Sezione di Bologna, viale Berti Pichat 6/2, I-40127 Bologna, Italy \label{inst11}
\and Department of Astrophysical Sciences, Princeton University, Princeton, NJ 08544, US \label{inst12}
\and Dipartimento di Fisica e Astronomia, Università di Bologna, via Gobetti 93 I-40127 Bologna, Italy \label{inst9}
\and Cogitamus Laboratory \label{inst19}
}

\date{Received 11 June 2020 / Accepted 30 July 2020}

\abstract{Chemical enrichment of the Universe at all scales is related to stellar winds and explosive supernovae phenomena. Metals produced by stars and later spread  throughout the intracluster medium (ICM) at the megaparsec scale become a fossil record of the chemical enrichment of the Universe and of the dynamical and feedback mechanisms determining their circulation. As demonstrated by the results of the soft X-ray spectrometer onboard \textsl{Hitomi}, high-resolution X-ray spectroscopy is the path to differentiating among the models that consider different metal-production mechanisms, predict the outcoming yields, and are a function of the nature, mass, and/or initial metallicity of their stellar progenitor. Transformational results shall be achieved through improvements in the energy resolution and effective area of X-ray observatories, allowing them to detect rarer metals (e.g. Na, Al) and constrain yet-uncertain abundances (e.g. C, Ne, Ca, Ni). The X-ray Integral Field Unit (X-IFU) instrument onboard the next-generation European X-ray observatory \textsl{Athena} is expected to deliver such breakthroughs. Starting from 100~ks of synthetic observations of 12 abundance ratios in the ICM of four simulated clusters, we demonstrate that the X-IFU will be capable of recovering the input chemical enrichment models at both low ($z = 0.1$) and high ($z = 1$) redshifts, while statistically excluding more than 99.5\% of all the other tested combinations of models. By fixing the enrichment models which provide the best fit to the simulated data, we also show that the X-IFU will constrain the slope of the stellar initial mass function within $\sim$12\%. These constraints will be key ingredients in our understanding of the chemical enrichment of the Universe and its evolution.
}

\keywords{X-rays: galaxies: clusters – galaxies: clusters: intracluster medium – galaxies: abundances – supernovae: general – galaxies: fundamental parameters – methods: numerical}

\titlerunning{Constraining the origin and models of chemical enrichment in galaxy clusters using the \textsl{Athena} X-IFU}
\authorrunning{F. Mernier et al.}

\maketitle

\section{Introduction}

The processes that lead to chemical enrichment of the Universe remain one of the major open questions in astrophysics. Most of the light elements (H, He, Li) were produced in the very first minutes of the Universe, during the primordial nucleosynthesis \citep{Cyburt2016BBN}. Metals (i.e. elements heavier than Li) are instead more recent, as most of this enrichment is related to supernovae events (SNe) and to stellar winds \citep{Burbidge1957Origin}. In fact, different types of stellar sources do not produce the same elements. Elements from O to Si are predominantly produced by fusion reactions in the outer shells of massive stars ($M \geq 10\,M_{\odot}$) during core-collapse supernovae (SN$_{\rm cc}$, see \citealt{Nomoto2013Nucleo} for a review), while heavier elements (i.e. Si to Fe) are mostly related to thermonuclear explosions of white dwarfs (WDs) -- former remnants of low-mass stars ($M \leq 8\,M_{\odot}$) -- that is, type-Ia supernovae (SN$_{\rm Ia}$, see \citealt{Maoz2014SNIa} for a review). Lighter elements (e.g. C, N, O) can also be related to radiative stellar winds when low-mass stars enter their asymptotic giant branch (AGB) phase \citep{Iben1983AGB}. 

Though understood to some extent, the physics of these mechanisms is not fully constrained. For instance, two main scenarios compete to explain SN$_{\rm Ia}$ events: the explosive C-burning onto a WD (triggering the SN$_{\rm Ia}$ explosion) may be ignited either by  (i) the accretion of matter from a companion star when the WD mass approaches the Chandrasekhar mass (single-degenerate model, \citealt{Whelan1973Binaries}),  or (ii) the merger with another WD far below its Chandrasekhar mass (double-degenerate model, \citealt{Webbink1984Mergers}). The nature of the WD explosion itself also remains  unclear \citep{Hillebrandt2013SNIa} and end-of-life models invoke either deflagration or delayed-detonation scenarios to explain observations \citep{Iwamoto1999DD}. Similarly, for SN$_{\rm cc}$, metal production depends on the initial mass (hence lifetime) and metallicity of the star, $Z_{\rm init}$, the estimation of which is challenging. Beyond the progenitor mass, chemical enrichment at all scales is strongly coupled to the initial mass function (IMF), i.e. the relative proportion of low- and high- mass stars that are born within a given single stellar population. In fact, different IMFs result in different (relative) numbers of AGBs, SN$_{\rm cc}$, and SN$_{\rm Ia}$, affecting not only the chemical properties of galaxies, but also their entire evolution and feedback \citep[for a review, see e.g.][]{Bastian2010}. However, whether the IMF is the same for all galaxies or constant with time is still an open question \citep[e.g.][]{DeMasi2019}, and therefore observational signatures of the IMF at galactic scales and beyond are valuable in this respect.

Measurements through X-ray spectroscopy of the intra-cluster medium (ICM) performed by missions such as \textsl{XMM-Newton}, \textsl{Chandra,} or \textsl{Suzaku} provide outstanding results in recovering the chemical composition of the ICM and in probing the enrichment of the largest scales of the Universe \citep[for recent reviews, see][]{Werner2008Review,Mernier2018Review}. In fact, the investigation of radial metallicity profiles in the outskirts of these systems \citep{Werner2013Enrich,Urban2017Metallicity} highlighted strong evidence of an early metal-production scenario ($z > 2$--$3$), which predates the formation of clusters \citep[][for a review on metallicity profiles in numerical simulations, see Biffi et al. \citeyear{Biffi2018Rev}]{Fabjan2010,McCarthy2010,Biffi2017Clusters,Biffi2018} and is likely contemporary to the stelliferous epoch of the Universe \citep{Madau2014enrich}. Although active galactic nucleus (AGN) feedback of the dominant galaxy can in principle induce significant central metallicity variations as a function of outflow and/or jet events \citep[e.g.][]{Gaspari20113D}, the remarkable similarity in the spatial distribution of SN$_{\rm Ia}$ and SN$_{\rm cc}$ products even within the central metallicity peaks suggests a similar `early-enrichment' scenario for clusters (and groups) cores \citep{Simionescu2009,Million2011,Mernier2017Radial}. Nevertheless, the exact diffusion and transport mechanisms of metals from stars to the interstellar medium and beyond remains an open question in both observations \citep[][]{Kirkpatrick2011,DeGrandi2016,Urdampilleta2019} and simulations \citep[e.g.][]{Schindler2008,Greif2009}.

Additional information can be derived from the abundance ratios measured in the ICM. For example, the Mn/Fe and Ni/Fe ratios are both sensitive to the electron capture rates during SN$_{\rm Ia}$ explosions, and  are therefore crucial indicators of their progenitor channels \citep{Seitenzahl2013Mn,Mernier2016II,Hitomi2017Enrich}, while ratios of lighter elements can in principle provide constraints on the IMF and the initial metallicity of the SN$_{\rm cc}$ progenitors \citep[e.g.][]{dePlaa2007Abundances,Mernier2016II}. By pushing current observatories to their limit, recent studies derived constraints on the relative fraction of SN events that effectively contribute to the ICM enrichment. These measurements showed that SN$_{\rm Ia}$ and SN$_{\rm cc}$ contribute relatively equally to the overall chemical enrichment in the ICM \citep{Werner2006Abund, dePlaa2006Abund,Bulbul2012, Mernier2016II,dePlaa2017Cheers}. The comprehensive study of \citet{Simionescu2019SXS} compiled the most accurate abundance measurements of the Perseus cluster (taken with the \textit{XMM-Newton} RGS and Hitomi SXS instruments) and compared them to state-of-the-art SN$_{\rm Ia}$ and SN$_{\rm cc}$ yield models. Their surprising conclusion is that no current set of models is able to reproduce all the observed abundance ratios at once. In particular, the measured Si/Ar ratio tends to be systematically overpredicted by models, even when taking calibration and atomic uncertainties into account. Whereas further improvement of stellar nucleosynthesis models is expected, the non-negligible systematic errors associated to these observations and the lack of highly sensitive, spatially resolved high-resolution spectroscopy prevents us from steering any considerable change in the paradigm \citep[see also][]{deGrandi09Abundances}. In fact, measurements from currently flying missions are performed with moderate collective area, either over the whole X-ray band (0.4--10~keV) with modest spectral resolution ($> 100$~ eV), or with higher resolution dispersive spectroscopy but over the low E band (0.3--2~keV) and without any spatial resolution, which considerably limits interpretations. 

As revealed by \textsl{Hitomi} SXS \citep{Takahashi2018Hitomi}, more accurate measurements of rare elements (e.g. Ne, Cr or Mn) will be key to constraining SN models. These steps forward are expected through the successor of the SXS, Resolve, on the X-Ray Imaging and Spectroscopy Mission (\textsl{XRISM}, \citealt{XRISM2020}), but definitive answers will require instruments such as the X-ray Integral Field Unit \citep[X-IFU,][]{Barret2016XIFU, Barret2018XIFU} which will fly onboard \textsl{Athena} \citep{Nandra2013Athena}. With more than $3000$ micro-calorimeter pixels, the X-IFU will provide high-resolution spectroscopy on the 0.2--12\,keV energy band (2.5\,eV FWHM energy resolution out to 7\,keV) over an equivalent diameter of $5^{\prime}$ with 5$^{\prime \prime}$ pixel size. This will  (i) allow us to measure for the first time abundances of rare elements \citep{Ettori2013Athena}, (ii) provide unprecedented constraints on the spatial distribution of metals through the ICM \citep[][hereafter, Paper~I]{Cucchetti2018Abun}, and (iii) allow us to further explore the chemical evolution of the ICM down to $z \sim 2$ \citep{Pointecouteau2013}. Other astrophysical questions on the ICM, specifically its level and distribution of turbulence \citep[e.g.][]{Roncarelli2018XIFU,Clerc2019Turb,Cucchetti2019Turb}, will also be explored by the X-IFU in unprecedented detail.

In this paper, we perform a first investigation of the capabilities of the X-IFU in constraining the sources of the ICM chemical enrichment and the associated stellar IMF in galaxy clusters. To do so, we use inputs from hydrodynamical cosmological simulations as already adopted in Paper~I. The methods are described in Sect.~\ref{sec:methods}. Using the large number of theoretical models at our disposal, we first demonstrate the capabilities of the X-IFU in recovering the underlying enrichment mechanisms assumed in the simulations (Sect.~\ref{sec:enr}). By fixing the best-fit models to the data, a second study is performed to analyse the capabilities of the instrument in constraining important parameters linked to the IMF, such as its slope, shape, and the high-mass cut-off (Sect.~\ref{sec:imf}). Results obtained in both cases are then discussed and future prospects are addressed (Sect.~\ref{sec:disc}). Finally, we conclude our findings in Sect.~\ref{sec:conclusions}.

Throughout this paper, we assume that metals showing emission lines in the X-ray band are produced by three independent astrophysical sources: AGBs, SN$_{\rm cc}$, and SN$_{\rm Ia}$. The abundances and their associated ratios (i.e. normalised to the Fe abundance) are given in the solar units of \citet[][]{Anders1989Solar}. Although other reference tables are available in the literature \citep[e.g. the proto-solar, meteoritic abundances of][]{Lodders2009abund}, the choice of units here has no impact on our results, as long as they are used self-consistently from the simulations to the mock data analysis. We also assume a $\Lambda$-CDM cosmology with $\Omega_{\rm M} = 0.24$,  $\Omega_{\rm b} = 0.04$, $H_0 = 72$\,km s$^{-1}$ Mpc$^{-1}$, $\sigma_8 = 0.8$ and $n_s = 0.96$ as used in the original hydrodynamical simulations \citep{Rasia2015Simu}. Unless stated otherwise, the quoted errors consider a 68\% confidence level.

\begin{table*} [t!]
\caption{Mean abundance ratios with respect to iron, obtained from the simulation input values (`i' in the second column) and the analysis of the mock X-IFU data (`o' in the second column) of the $z \sim 0.1$ simulated clusters of Paper~I (100~ks of mock exposure for each system, labelled as cool-core [CC] or non-cool-core [NCC]). All values are within $R_{500}$.
} 
\setlength{\tabcolsep}{10pt}
\centering 
\begin{tabular}{c c c c c c c} 
\hline\hline \\[-0.8em] 
Ratio & Input (i) / & Cluster 1 & Cluster 2 & Cluster 3 & Cluster 4 & Sample average\\[0.2em] 
 & observed (o) & (CC) & (NCC) & (CC) & (NCC) &  \\[0.2em] 
\hline\\[-0.8em] 

C/Fe & i & 0.50 & 0.51 & 0.49 & 0.50 & 0.50 \\
 & o & 0.48 $\pm$ 0.34 & 0.35 $\pm$ 0.24 & 0.43 $\pm$ 0.28 & 0.48 $\pm$ 0.31 & 0.39 $\pm$ 0.15 \\
N/Fe & i & 0.39 & 0.40 & 0.40 & 0.39 & 0.39 \\
 & o & 0.32 $\pm$ 0.12 & 0.40 $\pm$ 0.15 & 0.52 $\pm$ 0.30 & 0.57 $\pm$ 0.20 & 0.41$\pm$ 0.08 \\
O/Fe & i & 0.71 & 0.66 & 0.62 & 0.73 & 0.68 \\
 & o & 0.73 $\pm$ 0.04 & 0.68 $\pm$ 0.04 & 0.67 $\pm$ 0.06 & 0.69 $\pm$ 0.06 & 0.69 $\pm$ 0.02 \\
Ne/Fe & i & 0.11 & 0.11 & 0.11 & 0.12 & 0.12 \\
 & o & 0.10 $\pm$ 0.03 & 0.12 $\pm$ 0.04 & 0.14 $\pm$ 0.06 & 0.20 $\pm$ 0.06 & 0.12 $\pm$ 0.02 \\
Na/Fe & i & 0.12 & 0.12 & 0.13 & 0.13 & 0.13 \\
 & o & 0.17 $\pm$ 0.12 & 0.34 $\pm$ 0.24 & 0.20 $\pm$ 0.13 & 0.21 $\pm$ 0.17 & 0.21 $\pm$ 0.08 \\
Mg/Fe & i & 0.61 & 0.61 & 0.62 & 0.61 & 0.61 \\
 & o & 0.63 $\pm$ 0.05 & 0.58 $\pm$ 0.06 & 0.59 $\pm$ 0.08 & 0.53 $\pm$ 0.08 & 0.59 $\pm$ 0.03 \\
Al/Fe & i & 0.14 & 0.11 & 0.11 & 0.16 & 0.13 \\
 & o & 0.16 $\pm$ 0.08 & 0.18 $\pm$ 0.10 & 0.21 $\pm$ 0.18 & 0.20 $\pm$ 0.13 & 0.18 $\pm$ 0.06 \\
Si/Fe & i & 1.20 & 1.21 & 1.22 & 1.20 & 1.21 \\
 & o & 1.24 $\pm$ 0.06 & 1.18 $\pm$ 0.05 & 1.27 $\pm$ 0.08 & 1.27 $\pm$ 0.09 & 1.23 $\pm$ 0.03 \\
S/Fe & i & 1.30 & 1.28 & 1.25 & 1.22 & 1.26 \\
 & o & 1.33 $\pm$ 0.08 & 1.30 $\pm$ 0.08 & 1.26 $\pm$ 0.12 & 1.18 $\pm$ 0.12 & 1.28 $\pm$ 0.05 \\
Ar/Fe & i & 0.23 & 0.23 & 0.23 & 0.23 & 0.23 \\
 & o & 0.19 $\pm$ 0.09 & 0.22 $\pm$ 0.11 & 0.20 $\pm$ 0.15 & 0.30 $\pm$ 0.17 & 0.22 $\pm$ 0.06 \\
Ca/Fe & i & 0.93 & 0.94 & 0.94 & 0.93 & 0.93 \\
 & o & 0.94 $\pm$ 0.20 & 0.71 $\pm$ 0.16 & 0.71 $\pm$ 0.24 & 0.99 $\pm$ 0.25 & 0.82 $\pm$ 0.10 \\
Ni/Fe & i & 2.17 & 2.11 & 2.23 & 2.17 & 2.17 \\
 & o & 2.12 $\pm$ 0.20 & 1.91 $\pm$ 0.21 & 2.11 $\pm$ 0.29 & 2.12 $\pm$ 0.32 & 2.05 $\pm$ 0.12 \\
\hline \hline
\end{tabular}
\label{table:ratios} 
\end{table*}

\section{Methods and simulations}
\label{sec:methods}

\subsection{Simulation setup and mock data}\label{subsec:setup}

As current measurements are not adequate in terms of joint spatial and spectral resolution to build a representative toy model and perform a feasibility study of the afore-presented science case by X-IFU, this study relies on input numerical simulations, which are then converted into mock X-IFU datasets. The full method is described extensively in Paper~I (to which we refer the reader for more details). 

In summary, four Lagrangian regions are extracted from a parent large-scale cosmological simulation and re-simulated at higher resolution including the treatment of hydrodynamical processes with sub-grid physics models \citep[using the tree-PM smoothed particle hydrodynamics code \texttt{GADGET-3}; see][]{Rasia2015Simu}. Of these four systems, two are `cool-core' and two are `non-cool-core', and each class contains both a low- and a high-mass cluster. Each of these four clusters is traced back successively to $z=0.1$ and $z=1$. Metal enrichment by AGBs, SN$_{\rm cc}$, and SN$_{\rm Ia}$ is implemented following the approach of \citet{Tornatore2007Enrichment}, in which yields for H, He, C, N, O, Ne, Na, Mg, Al, Si, S, Ar, Ca, Fe, and Ni are injected and tracked during the clusters evolution \citep{Biffi2018,Truong2019Metals}. These elements are assumed to be produced by (i) AGBs, (ii) SN$_{\rm cc}$, and (iii) SN$_{\rm Ia}$ following the yields of, respectively, (i) \citet{karakas2010}, (ii) \citet{WoosleyWeaver1995} with updates from \citet{romano2010}, and (iii) \citet{Thielemann2003}. 

These yield models are listed in boldface in Tables~\ref{tab:app:agb}, \ref{tab:app:sncc}, and ~\ref{tab:app:snia}  of Appendix~\ref{app:sne}. They are used as input in our cosmological simulations to estimate the abundance of the different metal species produced during the evolution of the stellar component, assuming proper lifetimes \citep{Padovani1993} and depending on its metallicity and mass distributions. For the latter, the simulations assume a Chabrier IMF of the stellar population \citep[][see also Sect.~\ref{sec:imf}]{chabrier2003}. For this reason, most of the produced $\alpha$-elements are the result of a complex distribution of AGB and SN$_{\rm cc}$ initial metallicities \citep{Tornatore2007Enrichment}, from which we aim further to recover the dominant contribution only. X-ray photons ultimately emitted by these simulations are then projected and converted into event lists suitable for synthetic observations with the X-IFU for fixed exposure time of 100~ks. This is done using the end-to-end simulator SIXTE \citep{Dauser2019SIXTE}\footnote{\url{https://www.sternwarte.uni-erlangen.de/research/sixte/}} and assuming a \texttt{vvapec} model for the parameters of each emitting particle \citep[for a similar approach, see also][]{Roncarelli2018XIFU}. As SIXTE is the official simulator for the X-IFU, the up-to-date response files of the instrument are used. For each of the nearby ($z=0.1$) systems, we simulate seven adjacent X-IFU pointings in order to fully cover their $R_{500}$ limits.

We analyse the mock data following Paper~I. Specifically, the projected mock spectra are extracted within $R_{500}$ for each cluster, and are then fitted with XSPEC \citep{Arnaud1996XSPEC} within 0.2--12 keV,  using the same \texttt{vvapec} as in the input simulations for consistency (with the normalisation, temperature, redshift, and relevant abundances as free parameters). This approach
thus neglects the impact of the uncertainties in the atomic models, which could be ultimately a limiting factor at the sensitivity to be reached by the X-IFU (for further discussion on systematic uncertainties and our approach to dealing with them, see Sect.~\ref{subsec:constraints} and Sect.~\ref{subsec:limitations}). Other typical observational effects are taken into account. In particular, the implementation of the background relies on a modelling of the astrophysical foreground (consisting of the local hot bubble and the Milky Way hot gaseous content, modelled by an unabsorbed and absorbed thermal plasma, respectively), the cosmic X-ray background (unresolved AGNs, modelled by a power law), and the instrumental background (directly implemented within SIXTE).  More details about this procedure are given in Paper~I. The obtained best-fit abundance ratios are shown in Table \ref{table:ratios} for each individual cluster as well as for the sample average (which are used throughout this paper). They can be directly compared to their corresponding input values from the simulations described above (listed on the same table). Results for other parameters, in particular the temperature, are further detailed in Paper~I.

\subsection{Production yields and principle of the comparison}\label{subsec:formalism}

The origin of metals in the ICM can be traced by $K$ distinct stellar sources of enrichment $k$ (in our case, $K = 3$ and $k$ refers successively to AGBs, SN$_{\rm cc}$, and SN$_{\rm Ia}$). As such, the total number of atoms, $N_{X, \textrm{tot}}$, of a given chemical element $X$ produced over time can be expressed as \citep{Gastaldello2002Ratios}:
\begin{align}\label{eq:Nxtot}
N_{X, \textrm{tot}} = \sum_{k=1}^K n_k N_{X,k}\,,
\end{align}
where $n_k$ are multiplicative constants representing the total number of each source needed to obtain the observed enrichment (i.e. number of SNe or AGBs), and $N_{X,k}$ represents the number of atoms produced by each source. This number can easily  be related to the mass $M_{X,k}$ of a produced element as usually provided by yield models from the literature:
\begin{align}
N_{X,k} = \frac{M_{X,k}}{\mu_X}\,,
\end{align}
where $\mu_X$ is the atomic weight of the element $X$ (in the same units as $M_{X,k}$). Usually, however, AGBs and SN$_{\rm cc}$ yield models are provided for a specific progenitor mass $m$ and need to be integrated over a single stellar population with a given IMF. Such an integrated mass $M^{\text{int}}_{X, k}$, assuming a power-law IMF, can be written as:
\begin{align}\label{eq:IMF_integration}
M^{\text{int}}_{X, k} = \frac{\int_{M_{\rm low}}^{M_{\rm cut}} M_{X,k} (m) \ m^{\alpha} \ dm}{\int_{M_{\rm low}}^{M_{\rm cut}} m^{\alpha} \ dm}\,,
\end{align}
where $\alpha$ is the slope of the IMF, $M_{\rm low}$ is the lowest stellar mass assumed for each source of enrichment (depending on the yield models, though typically $\sim 1 M_\odot$ and $\sim 10 M_\odot$ for AGBs and SN$_{\rm cc}$, respectively\footnote{Below $\sim 1 M_\odot$, stellar mass losses are negligible compared to the AGB wind losses above that limit. Below $\sim 10 M_\odot$, the core of a star is not massive enough to collapse and trigger a SN$_{\rm cc}$ explosion \citep[e.g.][]{Nomoto2013Nucleo}.}), and $M_{\rm cut}$ is the upper mass cut-off for the integration (yet uncertain, though typically assumed to be $\sim 40 M_\odot$). In the input numerical simulations, the actual IMF is represented by a Chabrier-like piece-wise power-law function with  $M_{\rm cut} = 40 M_{\odot}$ and the following slopes: $\alpha = - 1.2$ for $M \in [0.1, 0.3] M_{\odot}$, $\alpha = - 1.8$ for $M \in [0.3, 1.3] M_{\odot}$, and $\alpha = - 2.3$ for $M \in [1.3, 40] M_{\odot}$. The Salpeter-like IMF (often used in previous literature as well as in this paper) follows the slope $\alpha = - 2.35$ and is very similar to the Chabrier-like IMF beyond $\sim 1 M_\odot$.

The respective `absolute' (i.e. relative to H) abundance $\mathcal{A}_{X,k}$ of a given element $X$ is defined as:
\begin{align}
\mathcal{A}_{X,k} \equiv \frac{N_{X,k}/N_{\textrm{H},k}}{\left(N_{X}/N_{\textrm{H}}\right)_\textrm{ref}} = \frac{N_{X,k}}{N_{\textrm{H},k} \, \mathcal{A}_{X, \textrm{ref}}}\,,
\end{align}
 where $\mathcal{A}_{X, \textrm{ref}} \equiv \left(N_{X}/N_{\textrm{H}}\right)_\textrm{ref}$ are the selected solar (or proto-solar) reference abundances \citep[in our case,][]{Anders1989Solar}. 

Given the above equations, the associated $X$/Fe abundance ratio (i.e. relative to Fe) can be expressed as:
\begin{align}
\mathcal{Y}_{X,k} \equiv \frac{\mathcal{A}_{X,k}}{\mathcal{A}_{\textrm{Fe},k}} = \frac{M_{X,k}}{M_{\textrm{Fe},k}} \, \frac{\mu_\textrm{Fe}}{\mu_X} \, \frac{\mathcal{A}_{\textrm{Fe}, \textrm{ref}}}{\mathcal{A}_{X, \textrm{ref}}}\,.
\end{align}
Considering now all the sources of enrichment, and similar to Eq.~\ref{eq:Nxtot}, the overall abundance ratio $\mathcal{Y}_{X, \textrm{tot}}$ can be written
\begin{align}
\mathcal{Y}_{X, \textrm{tot}} = \sum_{k=1}^K a_k \mathcal{Y}_{X,k}\,,
\end{align}
where $a_k$ is another multiplicative constant (linked to $n_k$ with appropriate normalisation).  By definition $\mathcal{Y}_{\rm Fe, tot}=1$.

The previous equations are valid in the case of an overall study of the abundances, which accounts for the \emph{total} number of atoms produced in stars, galaxies, and the ICM \citep{Matteucci2005Ratios}. However, X-ray observations provide a direct measurement of the ICM content in heavy elements. In addition, while the previous approach is rather insensitive to the spatial variation of metals within the ICM itself (due to, e.g. turbulence or diffusion -- as clusters within $R_{500}$ are considered as closed-boxes), it does not consider the circulation timescale of metals from stars into the ICM \citep[see also][]{deGrandi09Abundances}, nor the possibly different locations of AGB, SN$_{\rm cc}$, and SN$_{\rm Ia}$ due to the different lifetimes of their progenitors. It does not account either for any potential difference in the metal distribution (relative and spatial differences) between the ICM phase and the stellar phase. A usual caveat, assumed here, is to consider that the fractions $a_k$ derived from ICM studies are representative of the number of events enriching the \emph{ICM-only} system, rather than the total galaxy cluster system (i.e. stellar \emph{and} ICM phases). In this case, the previous formula can be applied treating $\mathcal{Y}_{X, k}$ as `effective' yields, which describe the fractions of the stellar sources enriching the hot gas (equivalent if and only if the metal distribution is the same in galaxies and the ICM, \citealt{Humphrey2006Ratios}). 

Using X-ray measurements of abundance ratios, $\tilde{\mathcal{Y}_X}$, the consistency of a theoretical prediction of abundance ratios can be tested at the ICM level by fitting a linear combination of the integrated yields that minimises
\begin{align}
\chi^2 = \sum_{X} \frac{(\tilde{\mathcal{Y}}_{X} - \mathcal{Y}_{X, \textrm{tot}})^2}{\sigma_{\textrm{stat}, X}^2}\,,
\label{eq:fit}
\end{align}
where the sum is performed over the total number of available elements $X$, and $\sigma_{\textrm{stat}}$ is the statistical error of the measurements  $\tilde{\mathcal{Y}_X}$ (our strategy to deal with potential systematic errors is explained further in Sect.~\ref{subsec:input}). The recovered values of $a_k/\sum_{k=1}^K a_k$  represent the corresponding fractions of each source $k$ at play in the ICM enrichment.

A generic way of quantifying the accuracy of a model in describing observations is to compute the reduced chi-squared of the fit, $\chi^2_{\rm red}$. If large values are obtained, the predicted yields are not likely to represent the observations of the chemical enrichment (at least at the ICM level) in a realistic way. These results can also be further refined by applying additional observational constraints on the ratio of SN$_{\rm Ia}$-to-SNe \citep[e.g.][]{Mernier2016II}. Current observational studies are limited in this comparison by the large (statistical and/or systematic) in the measured abundance ratios, and the limited number of elements observed \citep[e.g.][]{Mernier2015Abund,Mernier2016I}. As shown in Table~\ref{table:ratios}, X-IFU observations will provide a wealth of new constraints to discard or verify certain models. By comparing our mock data (Sect.~\ref{subsec:setup}) to a significant number of available models, in the following sections we derive which set of models matches the ratios in the most statistically accurate way, assessing whether it is ultimately possible to recover the models originally used in the input simulation (described in detail in \citealt{Tornatore2007Enrichment,Biffi2017Clusters}).

\section{Constraining chemical enrichment models with the X-IFU}
\label{sec:enr}

Besides the three distinct and independent sources of enrichment considered here -- AGB, SN$_{\rm cc}$, or SN$_{\rm Ia}$, we also assume that the bulk of the enrichment is completed at $z = 0.1$ and does not differ from local clusters, which is a fair hypothesis supported by numerical and observational results \citep[e.g.][]{Ettori2015evol,McDonald2016,Biffi2017Clusters,Urban2017Metallicity,Mantz2017Enrich,Biffi2018Rev,Liu2020}. Section~\ref{subsec:highz}  further extends the comparison to the case of $z = 1$ clusters.

In the following sections, nucleosynthesis yields computed for various models from the literature are fitted to the averaged $X$/Fe abundance ratios measured by the synthetic X-IFU observations. This is done by using the \texttt{abunfit} package in \texttt{python}\footnote{\url{https://github.com/mernier/abunfit}}, which solves the minimisation problem in Eq.~(\ref{eq:fit}) for $a_{\rm SN_{\rm Ia}}$, $a_{\rm SN_{\rm cc}}$, $a_{\rm AGB}$ (based on the approach of \citealt{Ettori2002Metals}, \citealt{Gastaldello2002Ratios}, \citealt{Werner2006Abund}, \citealt{dePlaa2007Abundances}, \citealt{Mernier2016II}, and \citealt{Simionescu2019SXS}). The fit is then performed under the constraint:
\begin{align}\label{eq:fits}
a_{\rm AGB} \mathcal{Y}_{\rm Fe, AGB} + a_{\rm SN_{\rm cc}} \mathcal{Y}_{\rm Fe, SN_{\rm cc}}  +  a_{\rm SN_{\rm Ia}} \mathcal{Y}_{\rm Fe, SN_{\rm Ia}} = 1.
\end{align} 

The full list of models (along with their relevant specifics) used for this comparison, and covering a significant fraction of the recent literature, is detailed in Appendix~\ref{app:sne}. Most AGB and SN$_{\rm cc}$ models are tested individually over the available range of initial metallicities $Z_{\rm init}$. Among the tested SN$_{\rm cc}$ models, we also include a set of hypernova and pair-instability SN predicted yields  in order to explore the reproducibility of the abundance pattern through an enrichment from very massive, metal-poor stars. To ease the comparison, throughout this section AGB and SN$_{\rm cc}$ yields are first integrated using the same Salpeter-like power-law IMF with a slope $\alpha=-2.35$ \citep{Salpeter1955IMF}. Although, formally, not exactly the same as the Chabrier-like\ IMF used in the input simulations, the Salpeter-like parametrisation is reasonable as a first approximation, is the simplest one to use, and is usually considered in previous similar studies. Other choices of the IMF (including Chabrier-like, which marginally improves our fit) are explored in Sect.~\ref{sec:imf}. A summary of our fits and methods is shown in Table~\ref{table:param}.

\begin{table*} [t!]
\caption{Summary of the fits performed on our mock observations throughout this study.
} 
\centering 
\begin{tabular}{c c l l c l} 
\hline\hline \\[-0.8em] 
Sect. & $z$ & Combination & Free parameters & D.o.f. & Total number of comb. \\[0.2em] 
\hline\\[-0.8em] 
\ref{subsec:input} & 0.1 & (AGB+SN$_{\rm cc}$+SN$_{\rm Ia}$)$_\text{input}$ & 2 (SN$_{\rm Ia}$/SNe, AGB/SNe) & 8 & $4\times5\times1 = 20$ \\
\ref{subsec:constraints} & 0.1 & AGB+SN$_{\rm cc}$+SN$_{\rm Ia}$ & 2 (SN$_{\rm Ia}$/SNe, AGB/SNe) & 8 & $4\times24\times182 = 17\,472$ \\
\ref{subsec:2snia} & 0.1 & AGB+SN$_{\rm cc}$+SN$_{\rm Ia,1}$+SN$_{\rm Ia,2}$ & 3 (SN$_{\rm Ia,1}$/SN$_{\rm Ia,1+2}$, SN$_{\rm Ia,1+2}$/SNe, AGB/SNe) & 7 & $4\times24\times78\times96 = 718\,848$ \\
\ref{subsec:highz} & 1 & (AGB+SN$_{\rm cc}$+SN$_{\rm Ia}$)$_\text{input}$ & 2 (SN$_{\rm Ia}$/SNe, AGB/SNe) & 6 & $4\times5\times1 = 20$ \\
\hline \hline
\end{tabular}
\label{table:param} 
\end{table*}

\subsection{Recovering the input simulated enrichment models}
\label{subsec:input}

As a first safety check, we aim to recover our X-IFU mock observed ratios with the combinations of the yield models that were considered as input in the simulated clusters. This corresponds to all the `K10+Ro10+Th03' combinations as listed in Appendix~\ref{app:sne} (4 AGB models $\times$ 5 SN$_{\rm cc}$ models $\times$ 1 SN$_{\rm Ia}$ model $ = 20$ combinations).

\begin{figure}[!t]
\centering
\includegraphics[width=0.49\textwidth]{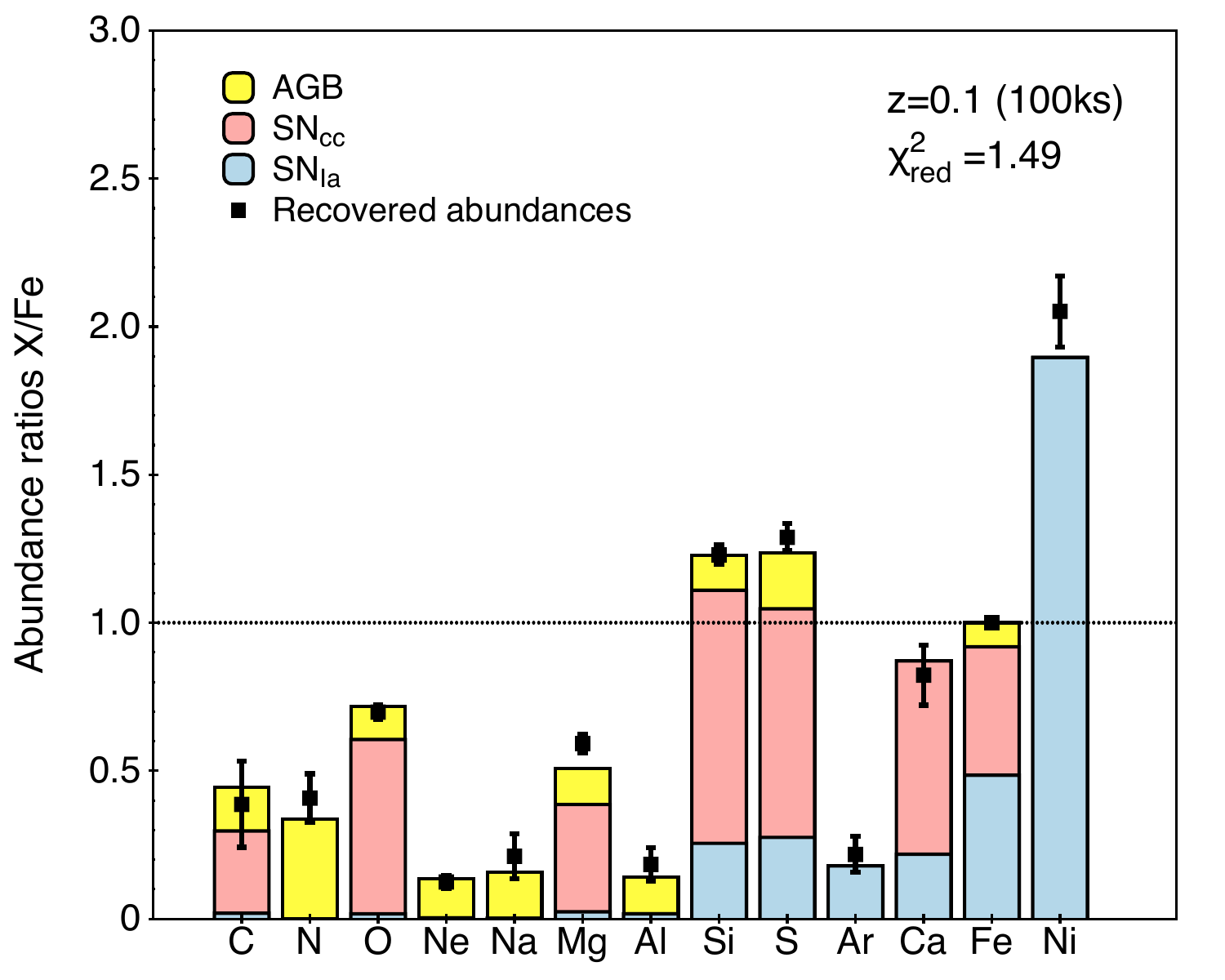}
\caption{Average abundance ratio $X$/Fe within $R_{500}$ over the cluster sample for $z = 0.1$ recovered using 100\,ks observations. These values are fitted using the enrichment yields derived from the references provided in Paper~I. The corresponding fitted contributions of SN$_{\text{Ia}}$ (blue), SN$_{\text{cc}}$ (magenta), and AGB stars (yellow) are shown as histograms.}
\label{fig:clusterratio1}
\end{figure}

Figure~\ref{fig:clusterratio1} shows the best-fit abundance pattern under these assumptions, obtained with $Z_{\rm init} = 0.01\,Z_{\odot}$ for SN$_{\rm cc}$ yields and $Z_{\rm init} = 1\,Z_{\odot}$ for AGB yields\footnote{As stellar evolution within our input simulations provided various initial metallicities for AGB and SN$_{\rm cc}$ progenitors, the best-fit $Z_{\rm init}$ values reported here translate their average contribution to the enrichment.}. With only two element ratios not perfectly recovered (Mg/Fe and Ni/Fe), the agreement between the mock ratios and best-fit models is good, leading to $\chi^2_\text{red} = 1.49$. This is particularly reassuring as under our assumptions it demonstrates the remarkable ability of the X-IFU to reproduce the {true} chemical composition of the ICM. We note that the exquisite precision achieved on most ratios naturally inflates $\chi^2_\text{red}$ for the elements whose yields are not perfectly recovered. In fact, when ignoring the Mg/Fe and Ni/Fe ratios, the fit further improves with $\chi^2_\text{red} = 0.55$. The precise reasons of the $>$1$\sigma$ discrepancies on these two specific ratios may be various (e.g. numerical issues, projection effects\footnote{As seen from Figs.~5 and 8 of Paper~I, the simulation-to-data discrepancies of the Mg/Fe ratio are mainly present in Cluster 3 (cool-core), while the other systems do not seem to be affected. While this may suggest that projection effects play a role in such simulations-to-data discrepancies (especially if Cluster 3 is affected by large-scale motions or substructures), the absence of such discrepancies for other ratios originating from the same source of enrichment (e.g. O/Fe) prevents us from excluding the other possibilities mentioned in the text.}, contribution from less dominant $Z_{\rm init}$ as initially implemented) and are left for investigation in future work.

\subsection{Constraints on various enrichment models}
\label{subsec:constraints}

Going one step further, though still assuming that each source of enrichment can be described by one single model, we aim to test whether the observed abundance pattern can be (mis-) interpreted with other combinations of AGB+SN$_{\rm cc}$+SN$_{\rm Ia}$ yield models available from the literature. This is particularly important in order to demonstrate the ability of the X-IFU to {constrain} relevant astrophysical parameters on the stellar population itself (e.g. average initial metallicity of SN$_{\rm cc}$ progenitors, favoured explosion channel of SN$_{\rm Ia}$).

\begin{figure*}[!ht]
\centering
\includegraphics[width=0.47\textwidth]{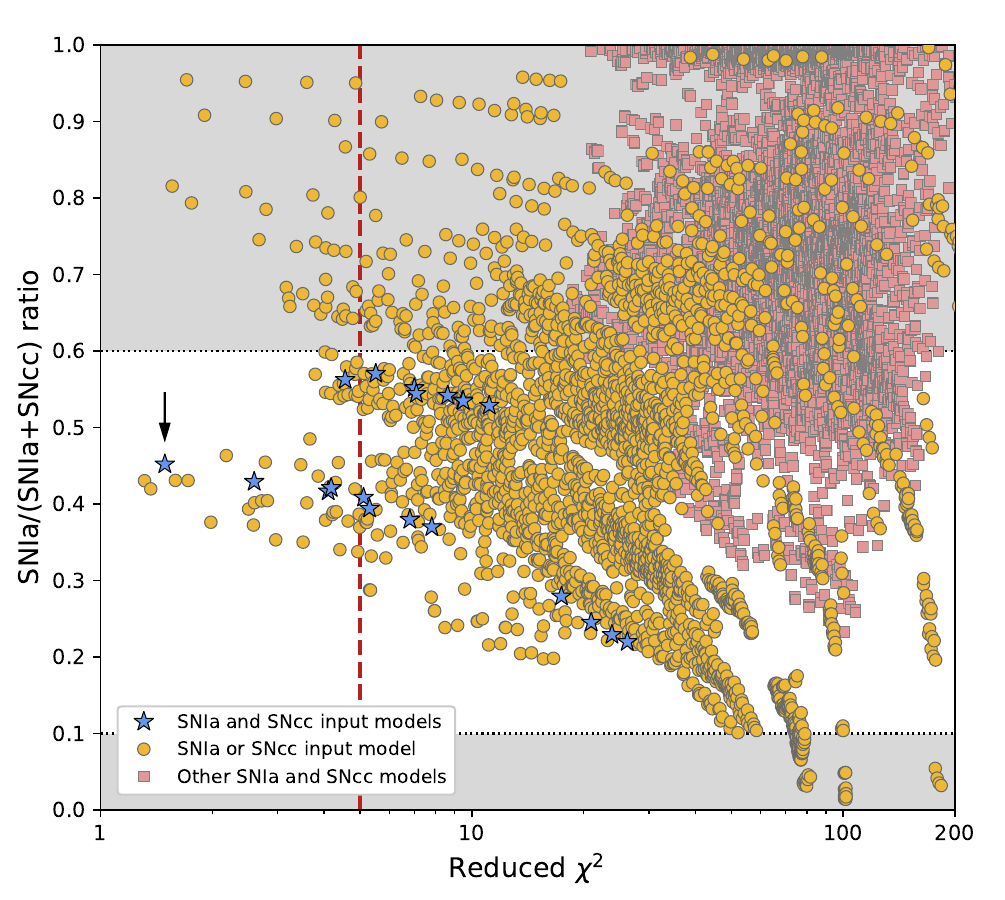}
\includegraphics[width=0.47\textwidth]{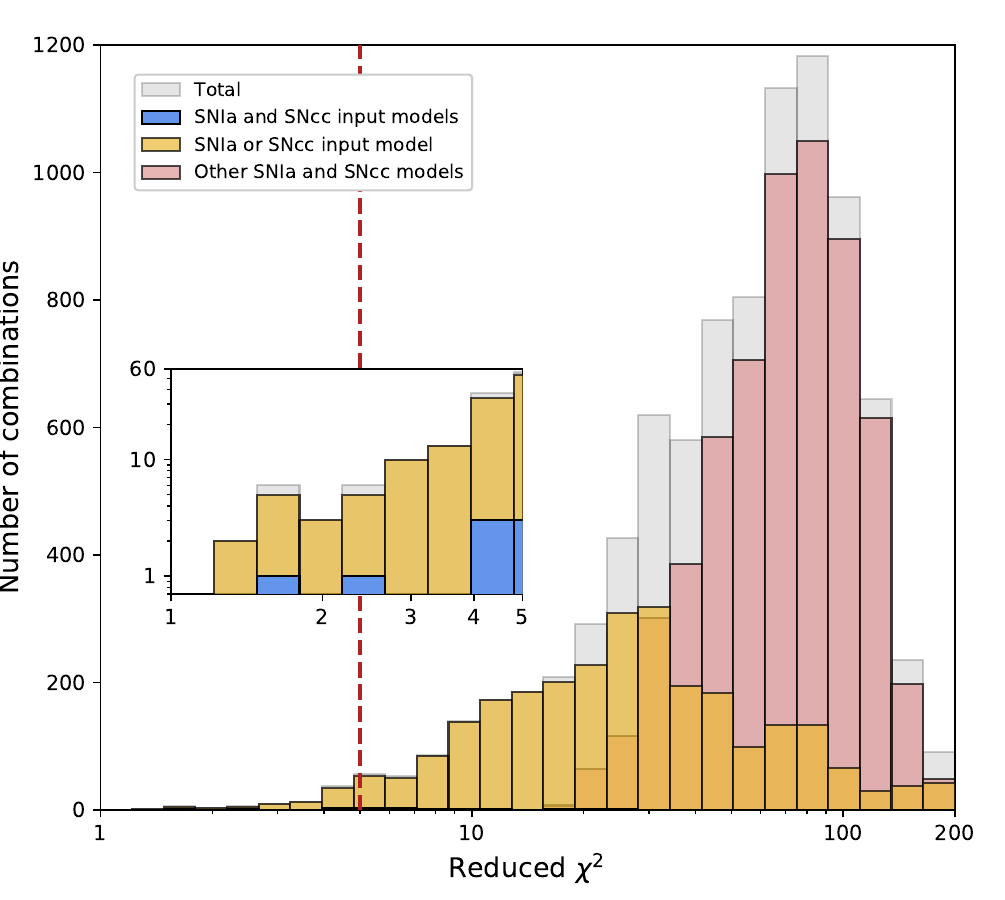}
\caption[Constraining supernovae models]{\textit{Left:} Best-fit $\chi_{\rm red}^{2}$ versus the SN$_{\rm Ia}$ fraction for all the possible combinations of (AGB+SN$_{\rm cc}$+SN$_{\rm Ia}$) yield models, fitted on the averaged X-IFU mock abundance ratios (see text). Combinations using (i) {both} SN$_{\rm cc}$ {and} SN$_{\rm Ia}$ models from the input cosmological simulations (blue stars), (ii) {one} of these models (for either SN$_{\rm cc}$ or SN$_{\rm Ia}$; yellow circles), and (iii) {neither} of these models (pink squares) are shown separately. The black arrow shows the `input' combination detailed in Fig.~\ref{fig:clusterratio1} and discussed in Sect.~\ref{subsec:input}. The white area delimitates realistic observational constraints previously obtained on the SN$_{\rm Ia}$ fraction contributing to the enrichment (see text). The dashed red line shows the `statistically acceptable' models ($\chi_{\rm red}^2 \leq 5$, see text). The list of best-fit models is provided in Table~\ref{tab:models}. \textit{Right:} $\chi_{\rm red}^{2}$ distribution of the same combinations. The same colour code as the left panel is applied.}
\label{fig:ratio_sne}
\end{figure*}

For each combination of models (i.e. a total of 4 AGB models $\times$ 24 SN$_{\rm cc}$ models $\times$ 182 SN$_{\rm Ia}$ models = 17\,472 combinations), we computed the reduced chi-squared of the fit with 8 degrees of freedom (corresponding to 12 elements - 3 models - 1). The corresponding $\chi^2_{\rm red}$ of each combination, along with the associated best-fit SN$_{\rm Ia}$ fraction -- i.e. SN$_{\rm Ia}$/SNe or, more formally, $a_{\rm SN_{\rm Ia}}/(a_{\rm SN_{\rm cc}}+a_{\rm SN_{\rm Ia}})$ -- is shown in Fig.~\ref{fig:ratio_sne} (left). All these combinations are also counted in histograms and their distribution is shown in Fig.~\ref{fig:ratio_sne} (right). We define a combination as `statistically acceptable' if it verifies  $\chi^2_{\rm red} \lesssim 5$. Such a conservative value is chosen arbitrarily to account for systematic error in the models or in the IMF and to avoid ruling out potentially significant models. The distribution indicates that, in the vast majority of cases ($17\,383$), the fit does not provide a good description of the mock measurements ($\chi_{\rm red}^2 > 5$). In fact we can distinguish two different features. A first peak ($\chi_{\rm red}^2 \sim 30$) corresponds to a fit in which \emph{one} out of the fitted models (either SN$_{\rm c}$ or SN$_{\rm Ia}$) is different from the input family of models used to perform the cosmological simulations. In that case, two of the models are close to reproducing correct metal production yields but the third one is not. Likewise, the second peak ($\chi_{\rm red}^2 \sim 80$) corresponds to the case where \emph{both} the SN$_{\rm cc}$ \emph{and} the SN$_{\rm Ia}$ models are different from the input family of models. In that case, none of them is consistent with the data set, thus providing a poorly accurate description of our mock abundance pattern.

\begin{table*} [!t]
\caption{Top ten best-fit combinations of AGB+SN$_{\rm cc}$+SN$_{\rm Ia}$ models with our X-IFU mock observed abundance ratios at $z = 0.1$, with $0.1 < \text{SN}_{\rm Ia}/\text{SNe} < 0.6$ (see text). The models mentioned here are all further detailed in Appendix~\ref{app:sne}. The models used as input in the cosmological simulations are given in bold.}
\setlength{\tabcolsep}{10pt}
\begin{center}
\begin{tabular}{c c c c c c} 
\hline \\[-0.8em] 
$\chi_{\rm red}^2$ & AGB  model & SN$_{\rm cc}$ model & SN$_{\rm Ia}$ model & SN$_{\rm Ia}$ frac. & AGB frac. \\[0.2em] 
\hline \hline\\[-0.8em] 

1.314 & \textbf{K10\_0.02} & \textbf{Ro10\_2E-3} & 100-3-c3 & 0.430& 0.340 \\
1.366 & \textbf{K10\_0.02} & \textbf{Ro10\_2E-3} & 500-5-c3 & 0.420 & 0.330 \\
1.490$^\dagger$ & \textbf{K10\_0.02} & \textbf{Ro10\_2E-4} & \textbf{Th03} & 0.452 & 0.319 \\
1.591 & \textbf{K10\_0.02} & \textbf{Ro10\_2E-3} & 100-5-c3 & 0.431 & 0.328 \\
1.722 & \textbf{K10\_0.02} & \textbf{Ro10\_2E-3} & 300-5-c3 & 0.431 & 0.325 \\
1.982 & \textbf{K10\_0.02} & \textbf{Ro10\_2E-3} & 500-3-c3 & 0.376 & 0.328 \\
2.182 & \textbf{K10\_0.02} & \textbf{Ro10\_2E-4} & 100-3-c3 & 0.463 & 0.337 \\
2.505 & \textbf{K10\_0.008} & \textbf{Ro10\_2E-3} & 500-5-c3 & 0.393 & 0.388 \\
2.583 & \textbf{K10\_0.02} & \textbf{Ro10\_2E-3} & 300-3-c3 & 0.372 & 0.325 \\
2.594 & \textbf{K10\_0.008} & \textbf{Ro10\_2E-4} & \textbf{Th03} & 0.429 & 0.377 \\
\hline
\end{tabular}
\end{center}
\label{tab:models} 
\tablefoot{$^\dagger$ This model combination is considered fixed in Sect.~\ref{sec:imf}.}
\end{table*}

From the list of models presented in Appendix~\ref{app:sne}, only 89 AGB+SN$_{\rm cc}$+SN$_{\rm Ia}$ combinations provide statistically acceptable results ($\chi^2_{\rm red} \lesssim 5$). This represents only 0.5\% of the total number of combinations. The X-IFU is therefore able to reject more than $\geq 99\%$ of the theoretical combinations of models tested here. A further refinement can be performed by analysing the results of the fit, in particular the relative contribution of each mechanism. From previous observations, we know that the observed SN$_{\rm Ia}$ fraction is, very conservatively, comprised between 0.1 and 0.6 \citep[see tables 5 and 6 in][]{deGrandi09Abundances}. As such, we can constrain the number of statistically accurate fits even further by requesting that $a_{\rm SN_{\rm Ia}}/(a_{\rm SN_{\rm Ia}}+a_{\rm SN_{\rm cc}})$ remains within these limits (Fig.~\ref{fig:ratio_sne}, left). Following this second selection, 53 combinations satisfy $\chi^2_{\rm red}\,<\,5$. In Table~\ref{tab:models}, we show the ten best-fit combinations satisfying this criterion.

While the combination of models used in the numerical simulations and recovered in Sect. \ref{subsec:input} is remarkably situated among the top of all these best-fit combinations (black arrow in Fig.~\ref{fig:ratio_sne}, left), we note that two other combinations \citep[using respectively the 100-3-c3 and 500-5-c3 SN$_{\rm Ia}$ models from][]{Leung2018sne} provide a slightly better $\chi^2_{\rm red}$. We discuss the reasons for this mis-interpretation in Sect.~\ref{subsec:ability} and further demonstrate that in more realistic conditions the X-IFU will be easily able to refine the discriminations between these combinations and strongly favour the one corresponding to the genuine sources of enrichment.

Among these statistically acceptable results, we also note that all the combinations predict correctly the input family of SN$_{\rm cc}$ models. In fact, the yellow circles of Fig.~\ref{fig:ratio_sne} (left) below the dashed red line are all combinations in which only the SN$_{\rm Ia}$ model was not consistent with the input simulations. This means that our fits are more inclined to recover an accurate SN$_{\rm cc}$ (family of) model(s) rather than an accurate SN$_{\rm Ia}$ model. However, the ability of the X-IFU to measure more ratios than presented here (Sect.~\ref{subsec:ability})  will eventually result in a much tighter filtering on the SN$_{\rm Ia}$ models.

Despite the conservative treatment of the systematic uncertainties considered above, one may wonder how our results would be altered if the X-IFU abundance measurements of one of the key elements turns out to be unreliable. To check this scenario, we re-fit our 17\,472 combinations ignoring successively one given $X$/Fe ratio. The main effect is a slight decrease of $\chi^2_{\rm red}$ without significantly altering the SN$_{\rm Ia}$ fraction, resulting in a horizontal left shift of the pattern seen in Fig.~\ref{fig:ratio_sne} (left). Ignoring Ni/Fe has the most noticeable impact, with a total of 443 combinations becoming statistically acceptable (i.e. $\sim$2.5\% of the total number of combinations). This is not surprising as, among all the ratios tested here, Ni/Fe is by far the most efficient at separating different SN$_{\rm Ia}$ models (see our discussion in Sect.~\ref{subsec:ability}). Ignoring other key ratios, such as Si/Fe or O/Fe (respectively 138 and 126 ``statistically acceptable'' combinations), has a much less pronounced impact on our results. Providing that the X-IFU will be able to measure the abundances of other key heavy elements (e.g. Cr, Mn; Sect.~\ref{subsec:ability}), our conclusions therefore hold even if not all ratios were considered as reliable.

\subsection{Lifting model degeneracies: the case of two SN$_{\rm Ia}$ models}
\label{subsec:2snia}

In theory, multiple sources of the same type may co-exist and enrich the ICM in comparable amounts. This might be notably the case for SN$_{\rm Ia}$, whose end of life can be significantly different depending on the single-degenerate versus double-degenerate scenario of their progenitors (and/or their deflagration vs. delayed-detonation thermonuclear explosions). Although this possibility is worth exploring, the use of multiple models to represent one class of physical events introduces one additional degree of freedom, which can be degenerate if the number and accuracy of the observed ratios is limited. In fact, this is one of the major limitations of current observatories when testing enrichment scenarios \citep{Mernier2016II}. To verify whether the accuracy expected from future X-IFU measurements is able to lift this degeneracy, we included an additional SN$_{\rm Ia}$ model to the fit to represent a more complex stellar reality. The same method as in Eq.~(\ref{eq:fits}) (Sect.~\ref{sec:enr}) is applied, considering two scalars, $a_{\rm SN_{\rm Ia}, 1}$ and $a_{\rm SN_{\rm Ia}, 2}$, in the fit. Specifically, one SN$_{\rm Ia}$ model is chosen to be near-Chandrasekhar (near-$M_\text{Ch}$, corresponding predominantly to the single-degenerate progenitor channel) while the other is chosen to be sub-Chandrasekhar (sub-$M_\text{Ch}$, corresponding predominantly to the double-degenerate progenitor channel; see also Table~\ref{tab:app:snia}). Assuming the same diversity of AGB and SN$_{\rm cc}$ models as in the previous section, this corresponds to a total of 4 AGB models $\times$ 24 SN$_{\rm cc}$ models $\times$ 78 SN$_{\rm Ia,1}$ models $\times$ 96 SN$_{\rm Ia,2}$ models = 718\,848 combinations of models. Although our input simulations include only \textit{one} (near-$M_\text{Ch}$) SN$_{\rm Ia}$ contribution (Sect.~\ref{subsec:setup}), this exercise is fully relevant as it shows whether or not the X-IFU measurement accuracies are good enough to avoid misinterpreting one dominant SN$_{\rm Ia}$ model with a combination of two SN$_{\rm Ia}$ models that would (incorrectly) contribute in comparable amounts to the enrichment.

We find that 472 of these AGB+SN$_{\rm cc}$+SN$_{\rm Ia, 1}$+SN$_{\rm Ia, 2}$ combinations offer a $\chi_{\rm red}^2$ value below 1.5 (13 of which contain one input SN$_{\rm Ia}$ model), and are thus better at reproducing our X-IFU mock abundance pattern than the input AGB+SN$_{\rm cc}$+SN$_{\rm Ia}$ set of models (Sect.~\ref{subsec:input}, Fig.~\ref{fig:clusterratio1}). Although this seems to be a large number, this represents only $\sim$0.07\% of the total number of our combinations including two SNIa models (near-$M_\text{Ch}$+sub-$M_\text{Ch}$). The more complex scenario where two SNIa models co-exist can be further constrained imposing a limit on the relative contributions from the two competing mechanisms. For instance, we can request the SN$_{\rm Ia, 1}$/SN$_{\rm Ia,2}$ ratio (namely, $a_{\rm SN_{\rm Ia}, 1}/a_{\rm SN_{\rm Ia}, 2}$) not to exceed one order of magnitude, that is, that it lies between 0.1 and 10. These limits are arbitrarily chosen to satisfy the condition that both mechanisms remain quantitatively comparable (otherwise a single model should suffice at first order). When applying this criterion, together with the same criterion as in Sect.~\ref{subsec:constraints} (i.e. $0.1 < \text{SN}_{\rm Ia 1+2}/\text{SNe} < 0.6$), the number of `realistic' combinations below $\chi_{\rm red}^2 < 1.5$ drops to 190 (i.e. $\sim$0.03\% of the total number of combinations). 

Admittedly, the present case includes more degeneracies than our previous, less complex attempts (i.e. simple AGB+SN$_{\rm cc}$+SN$_{\rm Ia}$ combinations). These numbers, and their interpretation as to the ability of the X-IFU to disentangle an ICM enrichment from one or two (or more) SN$_{\rm Ia}$ models, are further discussed in Sect.~\ref{subsec:ability}.

\subsection{Best fits and extension to higher redshift}
\label{subsec:highz}

The results shown above offer very promising perspectives for the X-IFU. Through simple linear fits and physical considerations, we are able to accurately recover the underlying enrichment model implemented in the hydrodynamical simulations. As shown in Fig.~\ref{fig:clusterratio1} and Sect.~\ref{subsec:input}, the best fit for the data set is obtained, at $z=0.1$, for $Z_{\rm init} = 0.01\,Z_{\odot}$ for SN$_{\rm cc}$ yields and $Z_{\rm init} = 1\,Z_{\odot}$ for AGB yields, with values of the reduced chi-squared of 1.49 for local clusters, nicely recovering the input yield models from the simulations. 

\begin{figure}[!t]
\centering
\includegraphics[width=0.49\textwidth]{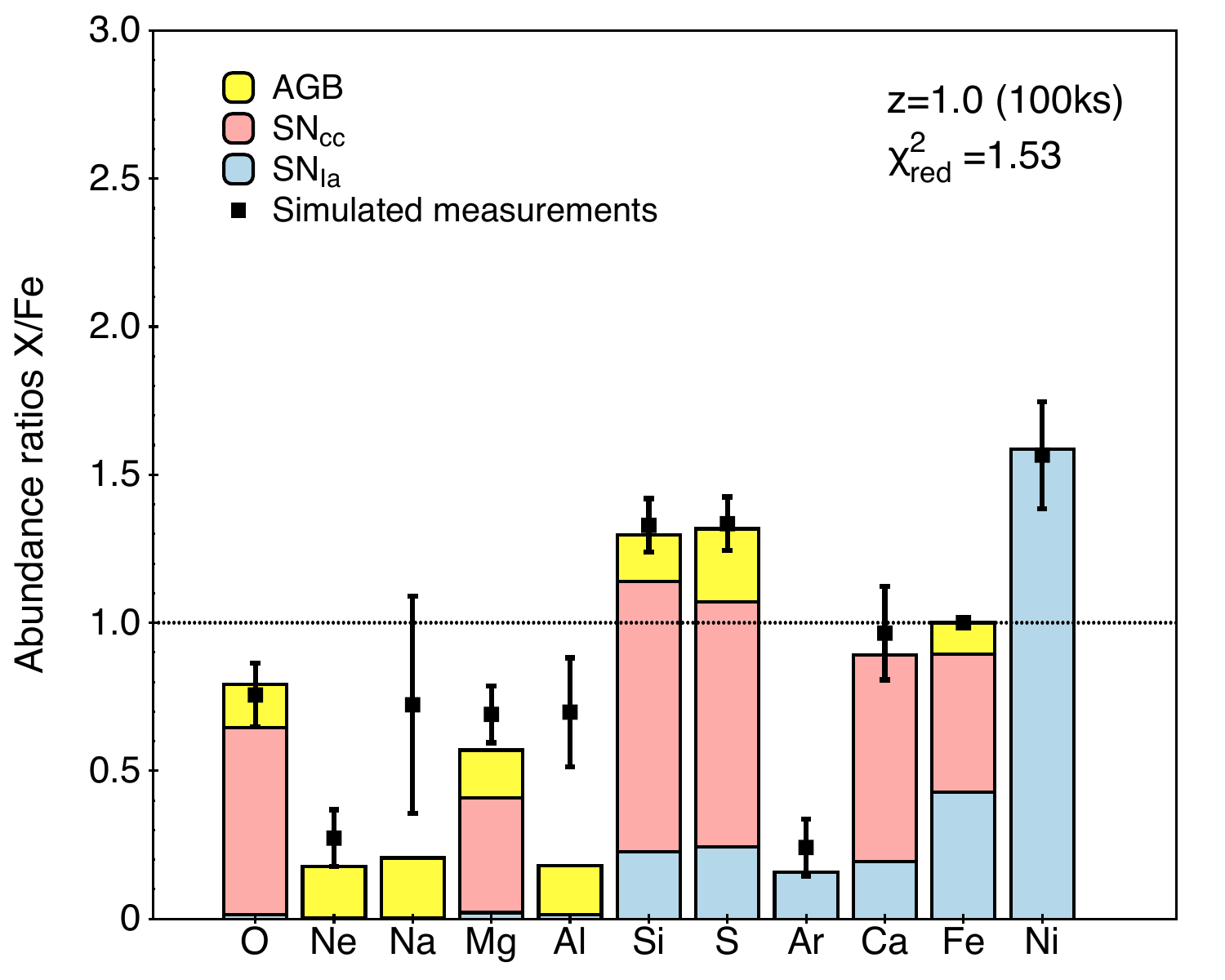}
\caption{Same as Fig.~\ref{fig:clusterratio1}, over the cluster sample at $z = 1$ (instead of  $z = 0.1$). Carbon and nitrogen are not shown as the lines are outside the energy bandpass of the X-IFU instrument.}
\label{fig:clusterratio2}
\end{figure}

The previous approach is also extended at $z = 1$ with excellent results  (see Fig.~\ref{fig:clusterratio2}). We find that the X-IFU is  still capable of accurately recovering abundance ratios within $R_{500}$ and the underlying input combination of models (same as Sect.~\ref{subsec:input}) with good accuracy ($\chi^2_{\rm red} = 1.53$). In addition, similarly to what has been reported in the input simulations \citep{Biffi2018}, our mock X-IFU observations consistently show no significant changes in the measured ratios between our two tested redshifts. 

We note that the aim of this exercise is limited to recovering the yields used in our input simulations. A full constraint among all the other considered models (as in Sect.~\ref{subsec:constraints} and Sect.~\ref{subsec:2snia}) is indeed less relevant at higher redshift if, as in our simulations, one assumes no substantial change in the physics of AGB, SN$_{\rm cc}$, and SN$_{\rm Ia}$ within the last 7-8 Gyr of ICM enrichment. Moreover, In the case of high-redshift systems, low-mass elements such as C and N can no longer be detected (as their emission lines fall outside the X-ray energy window) and measuring the abundance of rarer elements (e.g. Ne, Na, and Al) with sufficient statistical accuracy will require exposures deeper than 100~ks. At those high redshifts and for 100~ks exposures (or less), the derived AGB fraction is characterised by large uncertainties because the weight of those key low-mass elements becomes minimal in the fit. More accurate results call for better-adapted exposure strategies and larger sample studies in order to optimise the results for distant objects and to investigate the chemical enrichment of the AGB (in addition to SNe) across cosmic time.

\section{Recovering the IMF with the X-IFU}
\label{sec:imf}

\subsection{Effect of the IMF on yields}

Nucleosynthesis yields for AGB and SN$_{\rm cc}$ are usually given for a specific progenitor mass. As a consequence, they need to be integrated in a consistent (and realistic) way before fitting the result to our measurements (together with SN$_{\rm Ia}$ models). In the approach described above, in order to be consistent with the large majority of previous observational strategies \citep[][]{dePlaa2007Abundances,Mernier2016II,Simionescu2019SXS}, full integration of these yields is performed using a Salpeter-like power law. Though representative at first order, more advanced models can be used (e.g. Chabrier-like, as in our input simulations; \citealt{chabrier2003}) to provide a more accurate description of the low-mass parts of the IMF in the integration. Similarly, the mass cut-off beyond which a massive star directly collapses into a black hole (i.e. without ejecting freshly produced elements, and therefore not influencing AGB and SN$_{\rm cc}$ products) can be varied in the integration. Since all our models are integrated using the same IMF and mass cut-off (for consistency), no such related effects are expected in the above results. However, for a given AGB+SN$_{\rm cc}$+SN$_{\rm Ia}$ combination, one can vary $\alpha$ and $M_\text{cut}$ and study their effects on the fits. This is particularly relevant to verify whether (and to which extent) the X-IFU will be able to provide constraints on the shape, the slope, and the mass cut-off of the IMF.

\subsection{Recovering the IMF parameters for a fixed set of models}
\label{subsec:imf}

\begin{figure}
\centering
\includegraphics[width=0.49\textwidth, trim={0cm 0cm 0cm 0cm}]{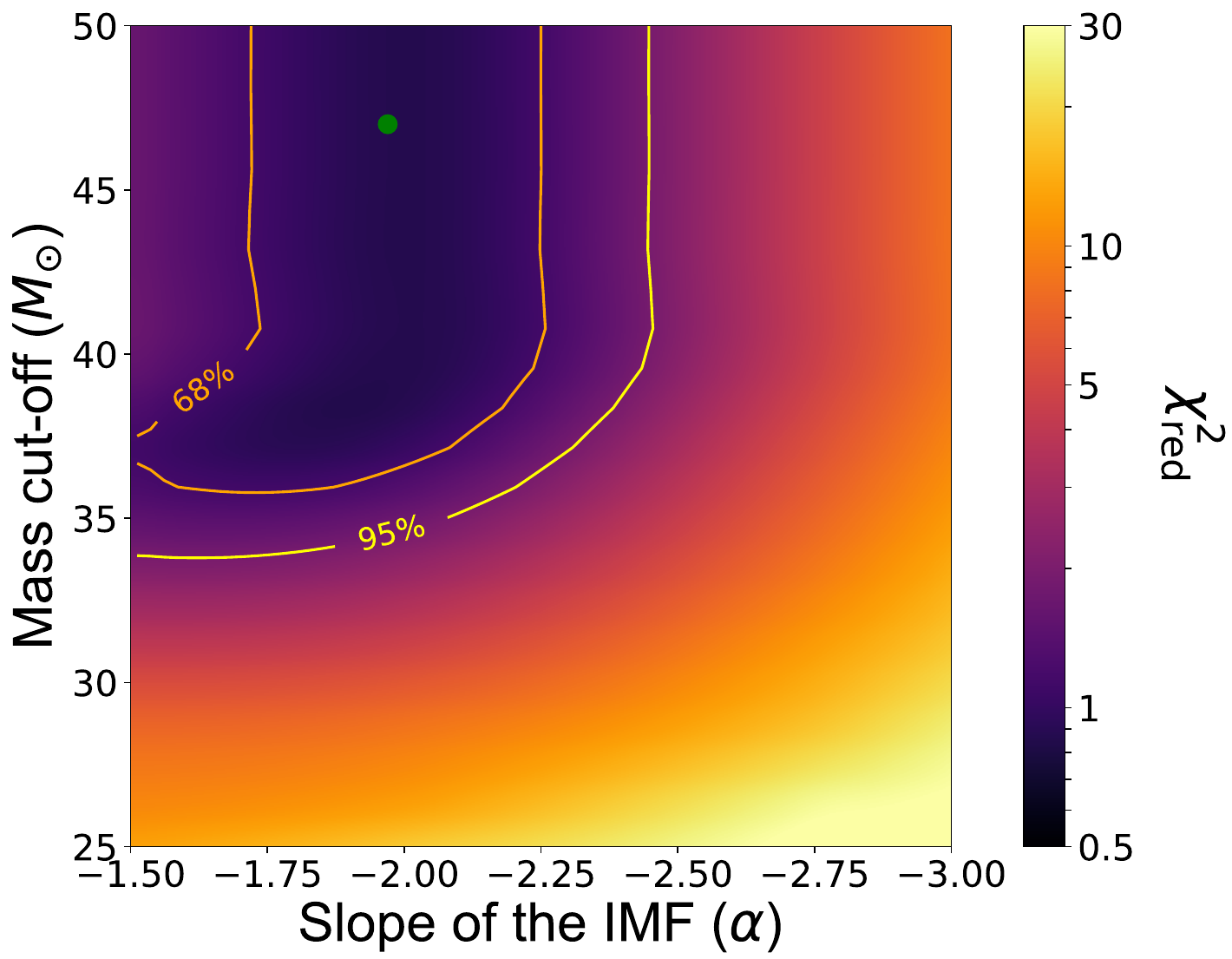}
\caption{Value of $\chi_{\rm red}^2$ of the fit of the data provided in Table~\ref{table:ratios}, with the best model from Table~\ref{tab:models}, for various Salpeter-like IMFs, as a function of the upper mass cut-off (in $M_{\odot}$) and slope of the power-law ($\alpha$). The contours delimitate the 68\% and 95\% confidence levels for the given degrees of freedom (in orange and yellow, respectively). The best-fit value is indicated by the green dot ($\alpha = -1.97$, $M_{\rm cut} = 47\, M_{\odot}$, $\chi_{\rm red}^2 = 0.87$).}
\label{fig:imf}
\end{figure}

The original IMF proposed by Salpeter assumes a value of $\alpha = - 2.35$ \citep{Salpeter1955IMF}. As our simulations initially assumed a Chabrier-like IMF, we aim to explore the effects of such changes on our best-fit combination of input yields (i.e. K10\_0.02+Ro10\_2E-4+Th03, see Sect.~\ref{subsec:input}) at $z = 0.1$. To do so, we re-integrate the total AGB and SN$_{\rm cc}$ yields over various arbitrary values of the IMF (Eq.~\ref{eq:IMF_integration}), with $\alpha \in [-1.5, -3.0]$, and using cut-off values of $M_{\rm cut} \in [25, 50]  M_{\odot}$. We then successively re-fit these modified models on our mock measurements. The results -- namely the variation of $\chi_{\rm red}^2$ on the $\{\alpha, M_{\rm cut} \}$ parameter plane -- are provided in Fig.~\ref{fig:imf}. For the given degrees of freedom, we notice that only a limited region of this plane provides results that are statistically consistent (i.e. within the 68\% and 95\% confidence intervals, corresponding respectively to $\Delta \chi^2$ values of 9.30 and 15.51 for 8 degrees of freedom) with our initial assumptions.

\begin{figure}
\centering
\includegraphics[width=0.49\textwidth, trim={0cm 0cm 0cm 0cm}]{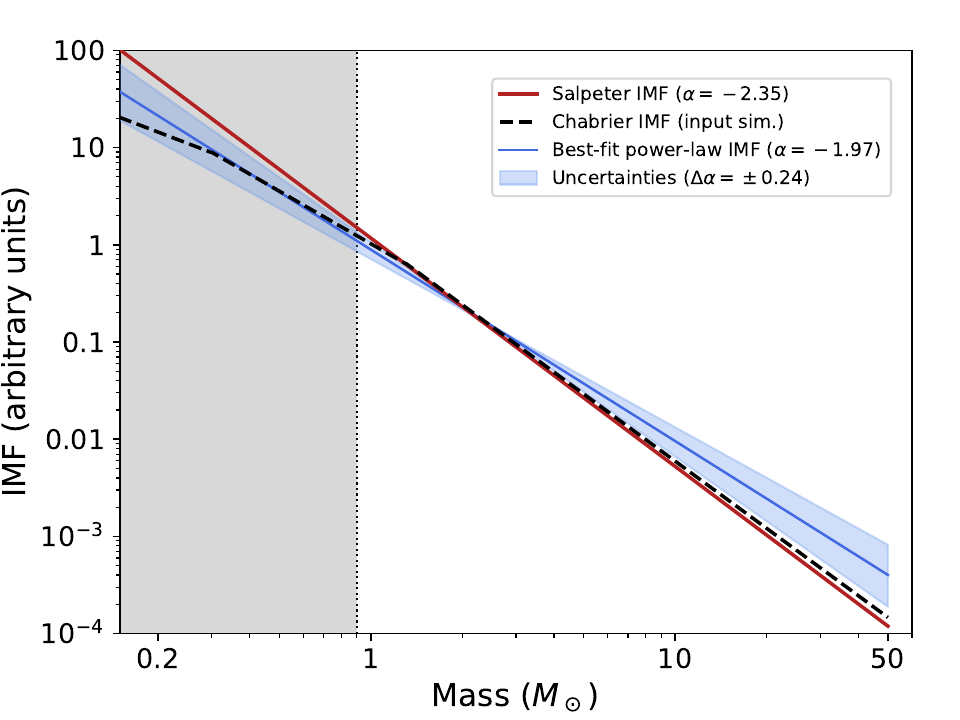}
\caption{Comparison of assumed slopes of the IMF throughout this study (normalised for their integral within [0.9--50] $M_\odot$ to be unity). The input Chabrier IMF (piece-wise function) as used in our input simulations is well reproduced by a power-law IMF of slope $\alpha = -1.97 \pm 0.24$ (see text). Stars with <0.9 $M_\odot$ are not expected to  directly enrich their surroundings, and therefore IMFs are not considered below that mass limit.}
\label{fig:imf_slopes}
\end{figure}

Clearly, the slope of the IMF has a significant effect on the consistency of the fit. Depending on its value, the reduced chi-square rises from $\sim 1$ to more than 10. The confidence interval contours show little dependence on the mass cut-off, such that the best-fit value is given by $\alpha = -1.97 \pm 0.24$ ($\pm 0.47$ at the 95\% level). Around these values, the mass cut-off of the IMF at low progenitor masses has a large effect on the yields. This is expected, as a low $M_{\rm cut}$ implies that an appreciable fraction of massive stars is not considered in the integration, thereby causing a bias toward low values some low-mass elements. When $M_{\rm cut} \geq 40 M_{\odot}$, the effect on the integration becomes negligible. This is related to the treatment of high-mass stars in the cosmological simulations \citep{Rasia2015Simu}, which are considered to collapse directly into black-holes beyond $40 M_{\odot}$, hence creating no chemical elements.

This best-fit value of the IMF slope ($\alpha \simeq -2$) can also be explained (Fig.~\ref{fig:imf_slopes}). When fitting the Chabrier-like piece-wise function (Sect.~\ref{subsec:formalism}) with a single slope power-law, the slope coefficient is indeed close to $\alpha \sim - 2$. Although yields from AGBs (and by extension SN$_{\rm cc}$) contribute only for progenitors above $0.9 M_{\odot}$, the same study as for the Salpeter-like IMF can in principle be performed also on Chabrier-like IMFs. When manually  re-integrating the baseline yields (Sect.~\ref{subsec:input}) with such a Chabrier-like IMF \citep[as initially assumed in the original hydrodynamical simulations;][]{Tornatore2007Enrichment}, we find slight improvements with respect to the best fits obtained so far (Sect.~\ref{subsec:input}; Table~\ref{tab:models}). In fact, $\chi_{\rm red}^2$ improves from 1.49 to 1.32 at $z=0.1$ (with the S/Fe and Ni/Fe ratios becoming $<$1$\sigma$ consistent with the yield predictions, as well as a slight improvement on the Mg/Fe ratio), and from 1.53 to 1.41 at $z=1$ (with improvements essentially on the Na/Fe and Al/Fe ratios). 

Using very simple considerations on the IMF, we showed that the X-IFU will be able to provide useful constraints on the (average) IMF of the stellar population(s) responsible for the enrichment. To some extent, it will be capable of distinguishing the mass cut-off of this function and to provide refinements on the value of the power slope (especially in the high-mass regime of SN$_{\rm cc}$ progenitors, where presumably the IMF is close to a single-slope power law). The observation of multiple clusters with the X-IFU and the accuracy of the recovered abundance ratios will provide an interesting tool for future IMF studies. Further discussion on this point and the ability of the X-IFU to favour different functional shapes of the IMF is provided in Sect.~\ref{subsec:discussion_IMF}.

\section{Discussion}
\label{sec:disc}

\subsection{Ability of the X-IFU to constrain nucleosynthesis models}
\label{subsec:ability}

In Sect.~\ref{subsec:input} and Sect.~\ref{subsec:constraints}, we show that the combination of AGB+SN$_{\rm cc}$+SN$_{\rm Ia}$ models representing the enrichment processes as injected in our cosmological simulations (blue stars in Fig.~\ref{fig:ratio_sne} left) are remarkably well recovered through the abundance ratios measured by the X-IFU. However, it also appeared that two other combinations (using SN$_{\rm Ia}$ models that were not initially considered in our simulations) provide a better fit to our mock abundance pattern. 

\begin{figure}[!]
\centering
\includegraphics[width=0.49\textwidth]{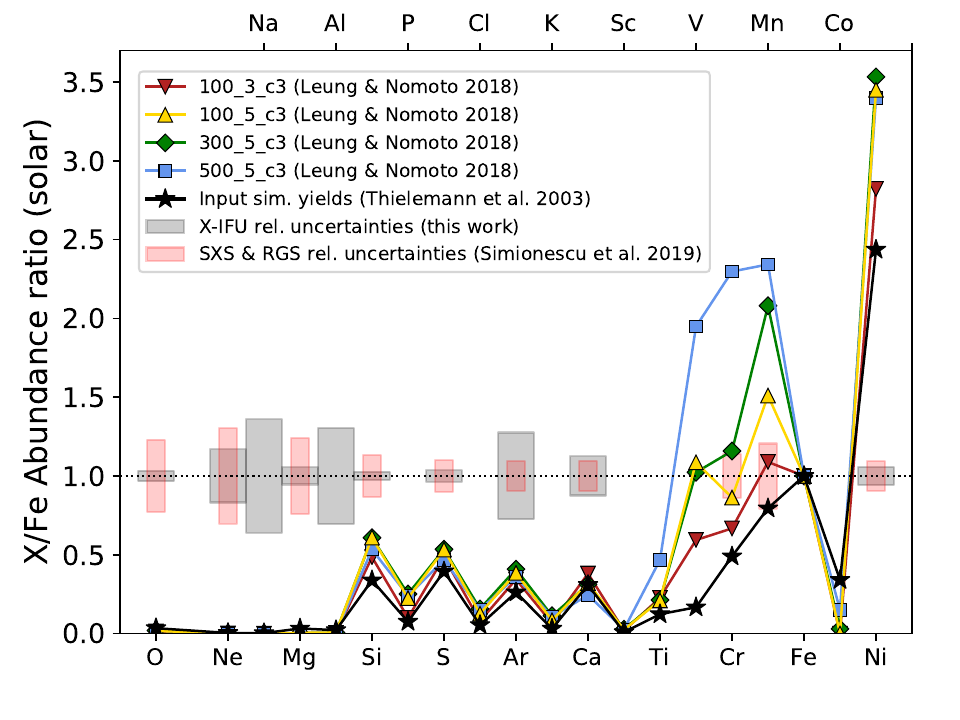}
\caption{Comparison of various $X$/Fe abundance ratios predicted by the SN$_{\rm Ia}$ yield models used in our top five `SN$_{\rm Ia}$ model' best fits (Table~\ref{tab:models}). Upper and lower horizontal axes mark respectively the odd- and even-Z elements. For comparison, we also show the {relative} (i.e. normalised to 1) observational uncertainties on the $X$/Fe ratios derived (i) in this work using the X-IFU (12 ratios, as tracked by our input simulations, C and N are not shown here), and (ii) using $\sim$300 ks of Hitomi/SXS (in synergy with $\sim$200 ks of \textit{XMM-Newton}/RGS) exposure \citep{Simionescu2019SXS}.}
\label{fig:SNIamodels}
\end{figure}

At first glance, this may appear as a source of concern regarding the ability of the X-IFU to correctly isolate the dominant physical and environmental mechanisms at play for stellar sources responsible for the enrichment. Nevertheless, one should keep in mind that in the present exercise we limit our study to the elements individually tracked within the numerical simulations, while the capabilities of X-IFU \citep[e.g.][]{Ettori2013Athena} extend to the abundance measurement of other Fe-peak elements as well, such as Ti, V, Cr, and Mn. Although these elements were not tracked in the original simulations \citep{Tornatore2007Enrichment}, they play a crucial role in the different SN$_{\rm Ia}$ explosion and progenitor scenarios. We illustrate this effect in Fig.~\ref{fig:SNIamodels}, where the predicted $X$/Fe ratios of the five most favoured SN$_{\rm Ia}$ models from our AGB+SN$_{\rm cc}$+SN$_{\rm Ia}$ fits (Table~\ref{tab:models}) are compared. It clearly appears that the Ti/Fe, V/Fe, Cr/Fe, and Mn/Fe ratios offer a powerful way to favour and/or rule out specific models. On the other hand, it is quite remarkable to note that, in this work, the (already impressive) ability of the X-IFU to efficiently favour our input model (namely, Th03) among 181 other ones from the literature was based solely on the Si/Fe, S/Fe, Ar/Fe, Ca/Fe, and Ni/Fe ratios, which individually exhibit limited model-to-model differences.

These additional constraints expected from Ti, V, Cr, and Mn will depend on the typical uncertainties that the X-IFU will measure for these abundances. While a complete re-run of cosmological simulations including these elements (along with a thorough mock spectral re-analysis) is out of the scope of this paper (see below), we highlight the remarkable constraints on Cr/Fe and Mn/Fe that were already achieved by the Hitomi SXS observations of the Perseus cluster \citep[][red areas in Fig.~\ref{fig:SNIamodels}]{Hitomi2017Enrich,Simionescu2019SXS}. Given the factor $\sim$10-12 improvement in weak-line sensitivity offered by the X-IFU in that energy band \citep{Barret2018XIFU} and the results presented in this work, it becomes clear that the X-IFU will ultimately be capable of isolating the most realistic physical and environmental constraints of SN progenitors responsible for the ICM enrichment.

The same reasoning could in principle be applied to the question of disentangling one from two SN$_{\rm Ia}$ models (Sect.~\ref{subsec:2snia}). Although the X-IFU is fully capable of favouring our input (single-SN$_{\rm Ia}$) enrichment model over more than 99.97\% of all the combinations including two SN$_{\rm Ia}$ (i.e. near-$M_\text{Ch}$+sub-$M_\text{Ch}$) models, an ideal goal would consist of isolating the former while disfavouring {all} the two-SN$_{\rm Ia}$ combinations. The inclusion of Cr/Fe and Mn/Fe ratios (and likely Ti/Fe and V/Fe) will certainly help to approach this limit. As mentioned earlier, this is particularly relevant given the high sensitivity of Mn to disentangle between near-$M_\text{Ch}$ and sub-$M_\text{Ch}$ models \citep[e.g.][]{Seitenzahl2013Mn,Mernier2016II,Hitomi2017Enrich}.

Admittedly, a complete demonstration of the X-IFU to favour one or two co-existing SN$_{\rm Ia}$ models would require not only the inclusion of these additional elements in the input simulations, but also the testing of the inverse exercise as proposed here -- i.e. inject \textit{two} SN$_{\rm Ia}$ progenitor channels in the simulations and test whether combinations with one SN$_{\rm Ia}$ model all fail to reproduce the resulting abundance pattern. Due to limited computing capacity, such re-runs are beyond the scope of this study and deeper investigation is therefore left to future work.

Another point of importance has been mentioned in \citet{Simionescu2019SXS}, where the authors show that no combination of the current models can
simultaneously reproduce all the measured abundance ratios in the Perseus cluster. This suggests that the current nucleosynthesis models need to be further improved to include more realistic physical processes that may have an impact on their ejected yields \citep[see also][]{deGrandi09Abundances}. Such improvements will be crucial for the era of exquisite ICM abundance accuracies that will be unveiled by XRISM and \textit{Athena}. They will also help to reduce the number of models that can be considered as physically realistic, further improving the constraints presented here.

\subsection{Constraints on the IMF}\label{subsec:discussion_IMF}

The investigation on the IMF performed here (Sect.~\ref{subsec:imf}) demonstrates the breadth of possibilities that will be accessible through the X-IFU. In fact, most of the possible caveats come essentially from our choice of the enrichment models rather than the capabilities of the instrument. Specifically, we notice that with slight changes in the IMF parameters, $\chi_{\rm red}^2$ can easily go above the previous threshold of five for our set of input models. The question of whether other sets of models could provide more accurate fits when integrated with slightly different IMFs thus naturally arises. However, integrating other models using modified IMFs (using the \texttt{abunfit} package)  confirms the results seen in Sect.~\ref{sec:imf}, that is, a best-fit slope typically around $\alpha \sim - 2$ and $M_{\rm cut} \sim 40 M_{\odot}$. This can be explained by the very simple shape of the IMF considered here, but could provide significantly different results with more complex functions. 

In fact, the IMF considered in the fits is simplistic, but ensures that we have only two parameters (slope and mass cut-off) to ease the interpretation. Other shapes can be considered, especially those with lower solar mass roll-offs (notably Chabrier-like, Fig.~\ref{fig:imf_slopes}), which assume a (realistic) finite number of stars at low masses. However, in this case, interpreting the contribution of each part of the IMF to the final integrated yields becomes more challenging. Furthermore, as most of these IMFs \citep[see a few examples in][]{chabrier2003} are approximated at high solar masses by a power law (similar to that of \citealt{Salpeter1955IMF}), no substantial change in the results is expected (mostly second-order changes are observed when using a Chabrier-like IMF). 

As shown in Sect.~\ref{subsec:imf}, the typical uncertainties on the ICM abundance ratios obtained with the 100~ks of X-IFU exposure on four nearby clusters will therefore allow us to derive constraints on the slope of the IMF of the order of $\sim$12\% (i.e. $\Delta \alpha = 0.24$). For comparison, while various measurements of the slope of the high-mass end of the IMF in the Galactic field, star-forming regions, associations, and star clusters show individual reported errors from $\sim$50\% down to a few percent, they scatter with deviations up to $\sim$60-70\% of the Salpeter value \citep[e.g. fig. 2 in][]{Bastian2010}. It becomes therefore evident that the \textit{Athena} mission will play a key role in substantially improving our understanding of the IMF, its global properties, and the question of its universality. Given that elliptical galaxies show hot atmospheres even in compact, isolated halos \citep[e.g.][]{Reiprich2002,Kim2019,Lakhchaura2019SMBH,Gaspari2019}, and under the condition that plasma emission codes will further provide the necessary accuracy, similar IMF studies can be expanded to the low-mass regime with similar statistics (e.g. lower luminosity is compensated by higher line emissivity of $\alpha$-elements due to cooler plasma temperature).

Because successive generations of stars become continuously enriched with metals (which may in turn affect their masses and lifetimes), it is also possible that the IMF evolves with cosmic time, with more weight toward massive stars at higher redshifts \citep[]{vanDokkum2008,Wang2011}. For instance, \citet{vanDokkum2008} reports a change of $\Delta \alpha \simeq 1.6$ between a sample of $0.02 < z < 0.83$ cluster galaxies and the canonical value around $1 M_\odot$. Although at a different mass range, this is larger than the typical uncertainty reported in Sect.~\ref{subsec:imf}. The question of how such a possible IMF evolution would affect the results presented here is not trivial, as it depends on (i)  exactly how the IMF evolves with time and (ii) at which exact redshift range  the metals seen today in the ICM were produced. While it is clear that good knowledge of the former will allow the X-IFU to bring constraints on the latter (or inversely), such constraints are difficult to quantify with the present simulations. Future work, with a time-changing IMF included in the input simulations, will certainly help to address this question.

As another consequence of stellar metallicities potentially evolving during the main epoch of enrichment, the simple assumption of using one $Z_{\rm init}$ in our models may limit the interpretation of the above results. While in such a case one can reasonably expect that $Z_{\rm init}$ represents the bulk of initial metallicities of SN$_{\rm cc}$ progenitors, the precise impact on the derived IMF is less clear. As shown in \citet{Mernier2016II}, the change of initial metallicity rather impacts the O/Ne ratio (see their fig. 2 right) while the effect of the IMF is rather reflected on the Ne/Mg and Ne/Si ratios (see their fig. 8). Although at first order we speculate that these effects are therefore limited, dedicated studies will be necessary to further quantify them.

\subsection{Possible improvements of the current approach}
\label{subsec:limitations}

From the above results, it is clear that the exquisite accuracy of the X-IFU in deriving ICM abundance ratios will allow us to favour and/or rule out various nucleosynthesis scenarios and conditions to explain the enrichment in galaxy clusters. This will considerably improve the constraints offered by the current missions \citep[e.g.][]{dePlaa2007Abundances,Mernier2016II,Mernier2018Ratios} as well as with the available Hitomi observations \citep[][]{Simionescu2019SXS}. As this work is the first attempt to quantify such constraints with the X-IFU, a few more points of discussion should be addressed, specifically in the context of improving our future observing strategies. 

First, and as outlined above, the obtained numbers (which are used to draw our conclusions) depend on the choice of the input yields in the hydrodynamical simulations used to generate synthetic X-IFU observations (described in Paper~I and references therein). Different results might be expected if for instance the yields assumed in the hydrodynamical simulations produce a truly unique chemical signature while on the contrary a large fraction of {all} the other model combinations happen to produce very similar abundance patterns. Although this is in practice unlikely to be the case here (see the rather large spread in the $\chi_{\rm red}^2$ of Fig.~\ref{fig:ratio_sne}), a proper validation would require tests with other input models, which we defer to future studies. We also note that no uncertainties are associated with these yields, simply because none are directly available in the current literature (although \citealt{deGrandi09Abundances} reported some yield uncertainties on the order of tens of percent). 

In this study, abundance ratios were derived within $R_{500}$ over four clusters. Although no strong observational evidence has been reported so far, the internal dynamical structure of clusters (e.g. turbulence, diffusion, AGN feedback) could to some extent affect the abundance ratios with radial and/or azimuthal inhomogeneities. This would naturally induce biases as a function of the spatial scale over which measurements are performed. However, this effect  can be safely neglected in our case. Indeed, Figs.~3, 4, and 5 of Paper~I clearly show that despite mild spatial metallicity inhomogeneities for a given system (and mild cluster-to-cluster differences in their metallicity profiles), the abundance {ratios} remain spatially uniform within uncertainties. It should be noted that, even though this picture is actually consistent with the absence of radial variation in abundance ratios reported on several observed systems \citep[e.g.][]{Ezer2017Abund,Mernier2017Radial} and in simulations \citep[][]{Biffi2017Clusters}, future X-IFU observations will be able to simultaneously measure abundance ratios integrated over large regions with very high accuracy (as shown in this paper) and investigate abundance ratios on much smaller (1D or 2D) scales. These local variations and anisotropies will also be key to constraining other important ICM observables, such as X-ray cavities and jets \citep[driven via AGN feedback; e.g.][]{Gaspari2020} or the halo structure \citep[shaped by mergers and sloshing; e.g.][]{Ettori2013Metals}.

Moreover, we cannot exclude the possibility that different elements do not enrich their surroundings with the same efficiency. For instance, specific elements freshly produced by SNe could be more (or less) easily depleted into dust than others before (or even after) ending up in the ICM. Multi-phase gas, which were not included in the present simulations due to sub-grid physics limitations, could also play an important role in this respect. The question of the interplay between hot gas and dust phases of metals (and their enrichment) in the central regions of galactic and cluster hot atmospheres has indeed only just begun to be explored \citep[e.g.][]{Panagoulia2015,Lakhchaura2019,Liu2019}. If this were the case, the observed abundance pattern should be reproduced with comprehensive chemo-dynamical models rather than linear fits of yield model combinations.

Despite our conservative approach to consider all fits with  $\chi^2_{\rm red} \le 5$ as statistically acceptable, a proper and thorough quantification of all the possible systematic uncertainties (e.g. instrumental response and calibration, spectral code uncertainties, background reproducibility, and subtle residual scatter in the ratios of different systems) that might affect future real observations will be important for the next steps of such studies. In high-redshift clusters, cosmic variance and AGN contamination might also require us to adapt the spatial and spectral analysis. While these systematic errors will contribute to increasing the total uncertainties, dedicated observing strategies (via e.g. larger samples or deeper exposures) will very likely help to compensate for this effect, in addition to optimising the ability of the X-IFU to constrain SN models. Future dedicated work on the total error budget, on the 12 abundance ratios studied here as well as additional ones (Sect.~\ref{subsec:ability}), will help to refine our results and predict the abilities of distinction between more subtle models. 

Finally, the approach adopted in this work relies on simple linear fits using $\chi^2$ statistics, allowing comparison with the previous observational studies -- using \textit{XMM-Newton} \citep[e.g.][]{Ettori2002Metals,dePlaa2007Abundances,Mernier2016II} and/or Hitomi \citep{Simionescu2019SXS}. Such a methodology will naturally tend to preferentially  reproduce ratios that have the smallest error bars. While this is not a problem for the context of this work, better approaches (e.g. using a Bayesian formalism) might be more appropriate for future real data, especially if some specific ratios suffer from additional systematic uncertainties (calibration, spectral codes, etc.) and/or if future nucleosynthesis calculations are provided with formal uncertainties on their predicted yields. A thorough, comprehensive comparison between different statistical methods -- listing advantages and drawbacks that are relevant for our goals -- is left for future work.

\section{Summary and conclusions}
\label{sec:conclusions}

In this paper, which naturally follows Paper~I, we use synthetic observations of clusters extracted from hydrodynamical simulations \citep{Rasia2015Simu} to demonstrate the capabilities of the X-IFU in constraining the chemical composition of the ICM. These mock observations were obtained using the X-IFU end-to-end simulator SIXTE \citep{Dauser2019SIXTE}. The measured abundance ratios were then compared to various combinations of existing AGB, SN$_{\rm cc}$ , and SN$_{\rm Ia}$ yields from the literature. Our main results can be summarised as follows.

\begin{enumerate}
\item The AGB+SN$_{\rm cc}$+SN$_{\rm Ia}$ combination of yield models that was used as input in the hydrodynamical simulations \citep{Tornatore2007Enrichment} is successfully recovered by the X-IFU, both at low ($z = 0.1$) and high ($z = 1$) redshifts. With $\chi_{\rm red}^2 = 1.49$ for local systems, this combination constitutes the third best fit (out of more than 17\,000 combinations) with our mock abundance ratios.
\item Complementary to the previous result, we show that a very large number of model combinations ($>$99.5\%) could be excluded as they provide a significantly worse fit. Expecting even further improvements when accounting for additional crucial ratios (e.g. Cr/Fe, Mn/Fe), this demonstrates that the X-IFU will be able to efficiently favour or rule out specific yield models, therefore providing valuable physical constraints on AGB, SN$_{\rm cc}$, and SN$_{\rm Ia}$ and their progenitors. These conclusions are essentially unchanged when the assumption of two co-exisiting SN$_{\rm Ia}$ models is considered.
\item For a fixed AGB+SN$_{\rm cc}$+SN$_{\rm Ia}$ model combination (i.e. the one that was recovered as input yields in the simulations), we tried to determine the possible constraints the X-IFU could provide on the stellar IMF, which is a critical parameter to understand the stellar and chemical evolution of the Universe. Through a simple integration of the yields using a Salpeter-like IMF \citep[i.e. power-law like, see][]{Salpeter1955IMF}, we demonstrate that the X-IFU will provide accurate values on the slope (within less than $\sim$12\%) and the upper mass cut-off of this function. Even further, the Chabrier-like shape of the IMF as used in our input simulations is recovered and favoured in the observations as well. Coupled with other observational evidence, the ability of the X-IFU to pick up a sensible IMF will help to better characterise it and test its (non-)universality.
\end{enumerate}

This study assesses -- for the first time -- the feasibility of a future instrument in providing constraints on the metal enrichment of the Universe by measuring the chemical composition of the ICM to an unprecedented level. Quite remarkably, most limitations are generally related to our current methodology (which should be considered as a first step) and/or model uncertainties rather than to the X-IFU capabilities. With multiple cluster observations at very high accuracy \citep[which will also help in estimating the total metal budget in the entire cluster volume; e.g.][]{Molendi2016Ab}, and provided that the upcoming XRISM observations will motivate continuous improvements on (i) nucleosynthesis calculations and (ii) spectral atomic codes \citep[as done already with the Hitomi observation of Perseus;][]{Hitomi2018Atomic}, the X-IFU will provide astronomers with new ways to investigate quantities such as metal production yields in stars, end-of-life models of compact objects, or the mass distribution of stars in the Universe.

\begin{acknowledgements}
The authors thank the anonymous referee for useful suggestions that helped to improve this paper, as well as A. Simionescu for insightful discussions. This work is supported by the Lend\"ulet LP2016-11 grant awarded by the Hungarian Academy of Sciences. V.B. acknowledges support by the DFG project nr. 415510302. V.B., S.B. and E.R. acknowledge financial contribution from the contract ASI-INAF n.2017-14-H.0. E.R. and S.B. acknowledge the ExaNeSt and Euro Exa projects, funded by the European Union's Horizon 2020 research and innovation programme, under grant agreement No 754337. S.B. also acknowledges support from the INFN INDARK grant. K.D. acknowledges support through ORIGINS, founded by the Deutsche Forschungsgemeinschaft (DFG, German Research Foundation) under Germany’s Excellence Strategy - EXC-2094 - 390783311. S.E. acknowledges financial contribution from the contracts ASI-INAF Athena 2015-046-R.0, ASI-INAF Athena 2019-27-HH.0, ``Attivit\`a di Studio per la comunit\`a scientifica di Astrofisica delle Alte Energie e Fisica Astroparticellare'' (Accordo Attuativo ASI-INAF n. 2017-14-H.0), and from INAF ``Call per interventi aggiuntivi a sostegno della ricerca di main stream di INAF''. M.G. is supported by the Lyman Spitzer Jr. Fellowship (Princeton University) and by NASA Chandra GO8-19104X/GO9-20114X and HST GO-15890.020-A grants. Our French and CNES colleagues are grateful to CNES for their outstanding support in developing the X-IFU for \textit{Athena}. SRON is supported financially by NWO, the Netherlands Organization for Scientific Research.
\end{acknowledgements}

\bibliography{paper}{}
\bibliographystyle{aa}

\newpage
\appendix

\onecolumn
\section{List of nucleosynthesis yield models}
\label{app:sne}

\begin{table*}[!h]
\centering
\caption{List of AGB yield models used in this work via the \texttt{abunfit} package (see below for references). All these models (bold) are used in our input cosmological simulations.}
\begin{tabular}{p{1.7cm} p{2.8cm} c p{11.0cm}}
\hline \\[-0.8em]
Category & Name & Ref. & Remarks \\[0.2em]
\hline  \hline \\[-0.8em]
\multicolumn{4}{c}{AGB} \\[0.2em] \hline \\[-0.8em]             

\textbf{Input}                   & \textbf{K10\_0.0001}  & $\pmb{\alpha} $ & $\pmb{Z_{\rm init} =  0.005 \, Z_{\odot}} $, $\pmb{M \in [0.9, 6]\, M_{\odot}} $ \\
                                                  & \textbf{K10\_0.004} & $\pmb{\alpha} $ & $\pmb{Z_{\rm init} = 0.2 \, Z_{\odot}} $, $\pmb{M \in [0.9, 6]\, M_{\odot}} $ \\
                                                  & \textbf{K10\_0.008} & $\pmb{\alpha} $ & $\pmb{Z_{\rm init} = 0.4 \, Z_{\odot}} $, $\pmb{M \in [0.9, 6]\, M_{\odot}} $ \\
                                                  & \textbf{K10\_0.02}  & $\pmb{\alpha} $ & $\pmb{Z_{\rm init} = 1 \,Z_{\odot}} $, $\pmb{M \in [0.9, 6]\, M_{\odot}} $ \\[0.2em] \\[-0.8em]  
\hline
\end{tabular}
\label{tab:app:agb}
\end{table*}

\begin{table*}[!h]
\centering
\caption{List of SN$_{\rm cc}$ yield models used in this work via the \texttt{abunfit} package (see below for references). Models used in our input cosmological simulations are given in bold.}
\begin{tabular}{p{1.7cm} p{2.8cm} c p{11.0cm}}
\hline \\[-0.8em]
Category & Name & Ref. & Remarks \\[0.2em]
\hline  \hline \\[-0.8em]
\multicolumn{4}{c}{SN$_{\rm cc}$} \\[0.2em] \hline \\[-0.8em]                                                   
\textbf{Input}                    & \textbf{Ro10\_0} & \textbf{a,b} & $\pmb{Z_{\rm init} = 0 \,Z_{\odot}}$, $\pmb{M \in [8, 50]\, M_{\odot}}$ \\
                                                  & \textbf{Ro10\_2E-6} &  \textbf{a,b} & $\pmb{Z_{\rm init} = 0.0001\,Z_{\odot}}$, $\pmb{M \in [8, 50]\, M_{\odot}}$ \\
                                                  & \textbf{Ro10\_2E-4} &  \textbf{a,b} & $\pmb{Z_{\rm init} = 0.01\,Z_{\odot}}$, $\pmb{M \in [8, 50]\, M_{\odot}}$ \\
                                                  & \textbf{Ro10\_2E-3} &  \textbf{a,b} & $\pmb{Z_{\rm init} = 0.1 \,Z_{\odot}}$, $\pmb{M \in [8, 50]\, M_{\odot}}$ \\
                                                  & \textbf{Ro10\_2E-2} &  \textbf{a,b} & $\pmb{Z_{\rm init} = 1\,Z_{\odot}}$, $\pmb{M \in [8, 50]\, M_{\odot}}$ \\[0.2em] \\[-0.8em]        

Nomoto                                   & No13\_SN$_{\rm cc}$\_0 & c &Core-collapse $Z_{\rm init} = 0 \,
                                                        Z_{\odot}$, $M \in [11, 140]\, M_{\odot}$ \\
                                                        & No13\_SN$_{\rm cc}$\_0.001 & c &Core-collapse $Z_{\rm init} = 0.05 \, Z_{\odot}$, $M \in [11, 40]\, M_{\odot}$ \\
                                                        & No13\_SN$_{\rm cc}$\_0.004 & c &Core-collapse $Z_{\rm init} = 0.2 \,Z_{\odot}$, $M \in [11, 40]\, M_{\odot}$ \\
                                                        & No13\_SN$_{\rm cc}$\_0.008 & c &Core-collapse $Z_{\rm init} = 0.4 \,Z_{\odot}$, $M \in [11, 40]\, M_{\odot}$ \\
                                                        & No13\_SN$_{\rm cc}$\_0.02 & c &Core-collapse $Z_{\rm init} = 1 \,Z_{\odot}$, $M \in [11, 40]\, M_{\odot}$ \\
                                                        & No13\_SN$_{\rm cc}$\_0.05 & c &Core-collapse $Z_{\rm init} = 2.5 \,Z_{\odot}$, $M \in [11, 40]\, M_{\odot}$ \\
                                                        & No13\_PISNe\_0 & c & Pair-instability SN$_{e}$ $Z_{\rm init} = 0 \,Z_{\odot}$, $M \in [140, 300]\, M_{\odot}$ \\
                                                        & No13\_SNe\_0 & c &Core-collapse $M \in [11, 140]\, M_{\odot}$ and pair-instability $M \in [140, 300]\, M_{\odot}$, $Z_{\rm init} = 0 \,Z_{\odot}$ \\
                                                        & No13\_HNe\_0 & c & Hyper-novae $Z_{\rm init} = 0 \,Z_{\odot} \,Z_{\odot}$, $M \in [20, 140]\, M_{\odot}$ \\
                                                        & No13\_HNe\_0.001 & c & Hyper-novae $Z_{\rm init} = 0.05 \,Z_{\odot}$, $M \in [20, 40]\, M_{\odot}$ \\
                                                        & No13\_HNe\_0.004 & c & Hyper-novae $Z_{\rm init} = 0.2 \,Z_{\odot}$, $M \in [20, 40]\, M_{\odot}$ \\
                                                        & No13\_HNe\_0.008 & c &Hyper-novae $Z_{\rm init} = 0.4 \,Z_{\odot}$, $M \in [20, 40]\, M_{\odot}$ \\
                                                        & No13\_HNe\_0.02 & c &Hyper-novae $Z_{\rm init} = 1 \,Z_{\odot}$, $M \in [20, 40]\, M_{\odot}$ \\
                                                        & No13\_HNe\_0.05 & c &Hyper-novae $Z_{\rm init} = 2.5 \,Z_{\odot}$, $M \in [20, 40]\, M_{\odot}$  \\[0.2em] \\[-0.8em]

Massive                                  & He0210\_SN$_{\rm cc}$\_0 & d,e & Core-collapse  $Z_{\rm init} = 0 \,Z_{\odot}$, $M \in [10, 100]\, M_{\odot}$ \\
                                                  & He0210\_PISNe\_0 & d,e & Pair-instability SN$_{e}$ $Z_{\rm init} = 0 \,Z_{\odot}$, $M \in [140, 260]\, M_{\odot}$ \\
                                                  & He0210\_SNe\_0 & d,e & Core-collapse $M \in [10, 100]\, M_{\odot}$ and pair-instability, $M \in [140, 260]\, M_{\odot}$, $Z_{\rm init} = 0 \, Z_{\odot}$  \\[0.2em] \\[-0.8em]  

Neutrino                                                 & Su16\_N20 & f & Incl. neutrino transport, calibrated for a \citet{Nomoto1988SNcc} progenitor to explode as SN1987A, $Z_{\rm init} = 1 \,Z_{\odot}$, $M \in [12, 120]\, M_{\odot}$ \\
                                                  & Su16\_W18 & f & Incl. neutrino transport, calibrated for a \citet{Utrobin2015SNcc} progenitor to explode as SN1987A, $Z_{\rm init} = 1 \,Z_{\odot}$, $M \in [12, 120]\, M_{\odot}$ \\[0.2em] \\[-0.8em]    
\hline
\end{tabular}
\label{tab:app:sncc}
\end{table*}

\begin{table}[!h]
\centering
\caption{List of SN$_{\rm Ia}$ yield models used in this work via the \texttt{abunfit} package (continues on next pages, see below for references). The model used in our input cosmological simulations is given in bold.}
\begin{tabular}{p{1.7cm} p{2.8cm} c p{11.0cm}}
\hline \\[-0.8em]
Category & Name & Ref. & Remarks \\[0.2em]
\hline  \hline \\[-0.8em]
\multicolumn{4}{c}{SN$_{\rm Ia}$} \\[0.2em] \hline \\[-0.8em]
\textbf{Input}                                                   & \textbf{Th03} & \textbf{1} & \textbf{1D deflagration} \\[0.2em] \\[-0.8em]

Bravo                                                    & DDTa & 2 & 1D delayed-detonation, fits the Tycho SNR, $\rho_{T,7}=3.9$ \\
                                                                & DDTb & 2 & (private comm., does not fit Tycho SNR, unpublished)  \\
                                                                & DDTc & 2 & 1D delayed-detonation, fits the Tycho SNR, $\rho_{T,7}=2.2$ \\
                                                                & DDTd & 2 & (private comm., does not fit Tycho SNR, unpublished)  \\
                                                                & DDTe & 2 & 1D delayed-detonation, fits the Tycho SNR, $\rho_{T,7}=1.3$ \\
                                                                & DDTf & 2 & (private comm., does not fit Tycho SNR, unpublished)   \\[0.2em] \\[-0.8em]

Ca-rich gap                             & CO.45HE.2 & 3 & Ca-rich SNe, $M_{\rm CO} =0.45$, $M_{\rm He} = 0.2$ \\
                                                & CO.55HE.2 & 3 & Ca-rich SNe, $M_{\rm CO} =0.55$, $M_{\rm He} = 0.2$\\
                                                & CO.5HE.15 & 3 &  Ca-rich SNe, $M_{\rm CO} =0.50$, $M_{\rm He} = 0.15$\\
                                                & CO.5HE.2 & 3 &Ca-rich SNe, $M_{\rm CO} =0.50$, $M_{\rm He} = 0.2$ \\
                                                & CO.5HE.2C.3 & 3 & Ca-rich SNe, $M_{\rm CO} =0.50$, $M_{\rm He} = 0.2$, 30\% mixing He-core layer\\
                                                & CO.5HE.2N.02 & 3 &Ca-rich SNe, $M_{\rm CO} =0.50$, $M_{\rm He} = 0.2$, 2\% N in He layer \\
                                                & CO.5HE.3 & 3 & Ca-rich SNe, $M_{\rm CO}= 0.50$, $M_{\rm He} = 0.3$\\
                                                 & CO.6HE.2 & 3 & Ca-rich SNe, $M_{\rm CO} =0.60$, $M_{\rm He} = 0.2$ \\[0.2em]  \\[-0.8em]       
                                                                                
2D                                                      & C-DEF & 4 & 2D deflagration $\rho_9 =2.9$  \\  
                                                        & C-DDT & 4 & 2D delayed-detonation $\rho_9 =2.9$, $\rho_{T,7}=1.0$ \\
                                                        & O-DDT & 4  & 2D delayed-detonation $\rho_9 =2.9$, $\rho_{T,7}=1.0$ , off-center ignition \\[0.2em]  \\[-0.8em]   

3D                                                & N1def & 5 & 3D deflagration $\rho_9 =2.9$, 1 ignition spot \\
                                                  & N3def & 5 & 3D deflagration $\rho_9 =2.9$, 3 ignition spots \\
                                                  & N5def & 5 & 3D deflagration $\rho_9 =2.9$, 5 ignition spots \\
                                                  & N10def & 5 & 3D deflagration $\rho_9 =2.9$, 10 ignition spots \\
                                                  & N20def & 5 & 3D deflagration $\rho_9 =2.9$, 20 ignition spots \\
                                                  & N40def & 5 & 3D deflagration $\rho_9 =2.9$, 40 ignition spots \\
                                                  & N100Hdef & 5 & 3D deflagration $\rho_9 =1.0$, 100 ignition spots \\
                                                  & N100def & 5 & 3D deflagration $\rho_9 =2.9$, 100 ignition spots \\
                                                  & N100Ldef & 5 & 3D deflagration $\rho_9 =5.5$, 100 ignition spots \\
                                                  & N150def & 5 & 3D deflagration $\rho_9 =2.9$, 150 ignition spots \\
                                                  & N200def & 5 & 3D deflagration $\rho_9 =2.9$, 200 ignition spots \\
                                                  & N300Cdef & 5 & 3D deflagration $\rho_9 =2.9$, 300 centred ignition spots \\
                                                  & N1600def & 5 & 3D deflagration $\rho_9 =2.9$, 1600 ignition spots \\
                                                  & N1600Cdef & 5 & 3D deflagration $\rho_9 =2.9$, 1600 centred ignition spots \\[0.2em]  \\[-0.8em]  \\[-0.8em]            
3D                                                & N1 & 6 & 3D delayed-detonation $\rho_9 =2.9$, 1 ignition spot \\
                                                  & N3 & 6 & 3D delayed-detonation $\rho_9 =2.9$, 3 ignition spots \\
                                                  & N5 & 6 & 3D delayed-detonation $\rho_9 =2.9$, 5 ignition spots \\
                                                  & N10 & 6 & 3D delayed-detonation $\rho_9 =2.9$, 10 ignition spots \\
                                                  & N20 & 6 & 3D delayed-detonation $\rho_9 =2.9$, 20 ignition spots \\
                                                  & N40 & 6 & 3D delayed-detonation $\rho_9 =2.9$, 40 ignition spots \\
                                                  & N100H & 6 & 3D delayed-detonation $\rho_9 =1.0$, 100 ignition spots \\
                                                  & N100 & 6 & 3D delayed-detonation $\rho_9 =2.9$, 100 ignition spots \\
                                                  & N100L & 6 & 3D delayed-detonation $\rho_9 =5.5$, 100 ignition spots \\
                                                  & N150 & 6 & 3D delayed-detonation $\rho_9 =2.9$, 150 ignition spots \\
                                                  & N200 & 6 & 3D delayed-detonation $\rho_9 =2.9$, 200 ignition spots \\
                                                  & N300C & 6 & 3D delayed-detonation $\rho_9 =2.9$, 300 centred ignition spots \\
                                                  & N1600 & 6 & 3D delayed-detonation $\rho_9 =2.9$, 1600 ignition spots \\
                                                  & N1600C & 6 & 3D delayed-detonation $\rho_9 =2.9$, 1600 centred ignition spots \\
                                                  & N100\_Z0.5 & 6 & 3D delayed-detonation $\rho_9 =2.9$, 100 ignition spots, $Z_{\rm init} = 0.5 \,Z_{\odot}$ \\
                                                  & N100\_Z0.1 & 6 & 3D delayed-detonation $\rho_9 =2.9$, 100 ignition spots, $Z_{\rm init} = 0.1 \,Z_{\odot}$  \\
                                                  & N100\_Z0.01 & 6 & 3D delayed-detonation $\rho_9 =2.9$, 100 ignition spots, $Z_{\rm init} = 0.01\, Z_{\odot}$  \\[0.2em]  \\[-0.8em]

\hline
\end{tabular}
\label{tab:app:snia}
\end{table}

\begin{table*}[!]
\centering
\begin{tabular}{p{1.7cm} p{2.8cm} c p{11.0cm}}
\hline \\[-0.8em]
Category & Name & Ref. & Remarks \\[0.2em]
\hline  \hline \\[-0.8em]

3D                                                      & N100\_c50 & 7 &  N100 with WD homogeneous core with 50\% C (mass) \\
                                                        & N100\_rpc20 & 7 & N100 with WD C-depleted core with 20\% C (mass)  \\
                                                        & N100\_rpc32 & 7 & N100 with WD C-depleted core with 32\% C (mass) \\
                                                        & N100\_rpc40 & 7 & N100 with WD C-depleted core with 40\% C (mass) \\[0.2em]  \\[-0.8em] 
        
LN18                                            & 050-1-c3-1P & 12 & 2D deflagr., $\rho_9 =0.5$, centred ignition, $Z_{\rm init} = 1\, Z_{\odot}$ \\              
                                                        & 100-1-c3-1P & 12 & 2D deflagr., $\rho_9 =1.0$, centred ignit., $Z_{\rm init} = 1\, Z_{\odot}$ \\
                                                        & 100-0-c3 & 12 & 2D del.-det., $\rho_9 =1.0$, centred ignit., $Z_{\rm init} = 0\, Z_{\odot}$ \\              
                                                        & 100-0.1-c3 & 12 & 2D del.-det., $\rho_9 =1.0$, centred ignit., $Z_{\rm init} = 0.1\, Z_{\odot}$  \\             
                                                        & 100-0.5-c3 & 12 & 2D del.-det., $\rho_9 =1.0$, centred ignit., $Z_{\rm init} = 0.5\, Z_{\odot}$ \\              
                                                        & 100-1-c3 & 12 & 2D del.-det., $\rho_9 =1.0$, centred ignit., $Z_{\rm init} = 1\, Z_{\odot}$ \\              
                                                        & 100-2-c3 & 12 & 2D del.-det., $\rho_9 =1.0$, centred ignit., $Z_{\rm init} = 2\, Z_{\odot}$ \\              
                                                        & 100-3-c3 & 12 & 2D del.-det., $\rho_9 =1.0$, centred ignit., $Z_{\rm init} = 3\, Z_{\odot}$ \\              
                                                        & 100-5-c3 & 12 & 2D del.-det., $\rho_9 =1.0$, centred ignit., $Z_{\rm init} = 5\, Z_{\odot}$ \\              
                                                        & 300-1-c3-1P & 12 & 2D deflagr., $\rho_9 =3.0$, centred ignit., $Z_{\rm init} = 1\, Z_{\odot}$ \\              
                                                        & 300-0-c3 & 12 & 2D del.-det., $\rho_9 =3.0$, centred ignit., $Z_{\rm init} = 0\, Z_{\odot}$ \\              
                                                        & 300-0.1-c3 & 12 & 2D del.-det., $\rho_9 =3.0$, centred ignit., $Z_{\rm init} = 0.1\, Z_{\odot}$  \\             
                                                        & 300-0.5-c3 & 12 & 2D del.-det., $\rho_9 =3.0$, centred ignit., $Z_{\rm init} = 0.5\, Z_{\odot}$  \\             
                                                        & 300-1-c3 & 12 & 2D del.-det., $\rho_9 =3.0$, centred ignit., $Z_{\rm init} = 1\, Z_{\odot}$ \\              
                                                        & 300-2-c3 & 12 & 2D del.-det., $\rho_9 =3.0$, centred ignit., $Z_{\rm init} = 2\, Z_{\odot}$ \\              
                                                        & 300-3-c3 & 12 & 2D del.-det., $\rho_9 =3.0$, centred ignit., $Z_{\rm init} = 3\, Z_{\odot}$ \\              
                                                        & 300-5-c3 & 12 & 2D del.-det., $\rho_9 =3.0$, centred ignit., $Z_{\rm init} = 5\, Z_{\odot}$ \\
                                                        & 500-1-c3-1P & 12 & 2D deflagr., $\rho_9 =5.0$, centred ignit., $Z_{\rm init} = 1\, Z_{\odot}$ \\
                                                        & 500-0-c3 & 12 & 2D del.-det., $\rho_9 =5.0$, centred ignit., $Z_{\rm init} = 0\, Z_{\odot}$ \\
                                                        & 500-0.1-c3 & 12 & 2D del.-det., $\rho_9 =5.0$, centred ignit., $Z_{\rm init} = 0.1\, Z_{\odot}$ \\
                                                        & 500-0.5-c3 & 12 & 2D del.-det., $\rho_9 =5.0$, centred ignit., $Z_{\rm init} = 0.5\, Z_{\odot}$ \\
                                                        & 500-1-c3 & 12 & 2D del.-det., $\rho_9 =5.0$, centred ignit., $Z_{\rm init} = 1\, Z_{\odot}$ \\
                                                        & 500-2-c3& 12 & 2D del.-det., $\rho_9 =5.0$, centred ignit., $Z_{\rm init} = 2\, Z_{\odot}$ \\
                                                        & 500-3-c3& 12 & 2D del.-det., $\rho_9 =5.0$, centred ignit., $Z_{\rm init} = 3\, Z_{\odot}$ \\              
                                                        & 500-5-c3& 12 & 2D del.-det., $\rho_9 =5.0$, centred ignit., $Z_{\rm init} = 5\, Z_{\odot}$ \\[0.2em]  \\[-0.8em]                   

HWD & N5\_hy & 8 & Hybrid WD (CO and ONe layers) \\[0.2em] \\[-0.8em]
        
GCD & GCD200 & 9 & Gravity-confined detonation \\[0.2em] \\[-0.8em]                                             
                        
DbleDet                                 & CSDD-L & 10 &  2D converging-shock double-deton.,  $M_{\rm CO} = 0.45 M_\odot$\\ 
                                                & CSDD-S & 10 & 2D converging-shock double-deton.,  $M_{\rm CO} = 0.58 M_\odot$ \\
                                                & ELDD-L & 10 & 2D edge-lit double-detonation,  $M_{\rm CO} = 0.45 M_\odot$\\
                                                & ELDD-S & 10 & 2D edge-lit double-detonation,  $M_{\rm CO}= 0.58 M_\odot$\\
                                                & HeD-L & 10 & 2D He detonation only,  $M_{\rm CO} = 0.45 M_\odot$\\
                                                & HeD-S & 10 & 2D He detonation only,  $M_{\rm CO} = 0.58 M_\odot$\\[0.2em]  \\[-0.8em]
                                                                                
ONe                                             & CO15e7 & 11 & 2D (sub-$M_\text{Ch}$) detonation carbon-oxygen WD, $\rho_9 =0.15$\\
                                                & ONe10e7 & 11 & 2D (sub-$M_\text{Ch}$) detonation oxygen-neon WD, $\rho_9 =0.10$\\
                                                & ONe13e7 & 11 & 2D (sub-$M_\text{Ch}$) detonation oxygen-neon WD, $\rho_9 =0.13$\\
                                                & ONe15e7 & 11 & 2D (sub-$M_\text{Ch}$) detonation oxygen-neon WD, $\rho_9 =0.15$\\
                                                & ONe17e7 & 11 & 2D (sub-$M_\text{Ch}$) detonation oxygen-neon WD, $\rho_9 =0.17$\\
                                                & ONe20e7 & 11 & 2D (sub-$M_\text{Ch}$) detonation oxygen-neon WD, $\rho_9 =0.20$\\[0.2em]  \\[-0.8em]  
                                                
Det                                             & det\_0.81 & 13 & 1D (sub-$M_\text{Ch}$) pure detonation, $M_\text{CO} = 0.81 M_\odot$, $\rho_7 =1.0$  \\
                                                & det\_0.88 & 13 & 1D (sub-$M_\text{Ch}$) pure detonation, $M_\text{CO} = 0.88 M_\odot$, $\rho_7 =1.45$  \\
                                                & det\_0.97 & 13 & 1D (sub-$M_\text{Ch}$) pure detonation, $M_\text{CO} = 0.97 M_\odot$, $\rho_7 =2.4$  \\
                                                & det\_1.06 & 13 & 1D (sub-$M_\text{Ch}$) pure detonation, $M_\text{CO} = 1.06 M_\odot$, $\rho_7 =4.15$  \\
                                                & det\_1.15 & 13 & 1D (sub-$M_\text{Ch}$) pure detonation, $M_\text{CO} = 1.15 M_\odot$, $\rho_7 =7.9$  \\
                                                & det\_1.06\_0.075Ne & 13 & 1D (sub-$M_\text{Ch}$) pure detonation, $M_\text{CO} = 1.06 M_\odot$, $\rho_7 =4.15$,  \\
                                                &  &  & C/O/Ne mass fraction = 0.425/0.5/0.075
                                                \\[0.2em]  \\[-0.8em]   
                                                
6D                                                & Sh18\_Ma\_b\_Zc\_d & 14 & 3D (sub-$M_\text{Ch}$) dynamically-driven double-degenerate double detonation, 159 models in total for different WD masses ($\text{Ma} \in \{0.8, 0.85, 0.9, 1.0, 1.1\}\,M_{\odot}$), C/O compositions ($\text{b} \in \{30/70, 50/50\}$), metallicity $Z_{\rm init}$ ($\text{Zc} \in \{0, 0.005, 0.01, 0.02\}$), and normalizations of the $^{12}$C+$^{16}$O reaction rate ($\text{d} \in \{0.1, 1.0\}$)\\[0.2em] \\[-0.8em]             
                                                                
Merger                                    & 09\_09 & 15 & 3D (sub-$M_\text{Ch}$) violent WD merger (double-degen.), 0.9+0.9 $M_\odot$ \\
                                                  & 11\_09 & 16 & 3D (sub-$M_\text{Ch}$) violent WD merger (double-degen.), 1.1+0.9 $M_\odot$ \\[0.2em] \\[-0.8em]       

\hline
\end{tabular}
\end{table*}

\begin{table*}[!]
\centering
\begin{tabular}{p{1.7cm} p{2.8cm} c p{11.0cm}}
\hline \\[-0.8em]
Category & Name & Ref. & Remarks \\[0.2em]
\hline  \hline \\[-0.8em]
Merger2                                   & 09\_076 & 17 & 3D (sub-$M_\text{Ch}$) violent WD merger (double-degen.), 0.9+0.76 $M_\odot$, $Z_{\rm init} = 1\, Z_{\odot}$ \\
                                                  & 09\_076\_Z0.01 & 17 & 3D (sub-$M_\text{Ch}$) violent WD merger (double-degen.), 0.9+0.76 $M_\odot$, $Z_{\rm init} = 0.01\, Z_{\odot}$ \\[0.2em] \\[-0.8em]  

\hline
\end{tabular}
\tablebib{\\ 
SN$_{\rm Ia}$: (1) \citet{Thielemann2003}; (2) \citet{Badenes2006SN}; (3) \citet{Waldman2011sne}; (4) \citet{Maeda2010sne}; (5) \citet{Fink2014sne}; (6) \citet{Seitenzahl2013DD}; (7) \citet{Ohlmann2014sne}; (8) \citet{Kromer2015sne}; (9) \citet{Seitenzahl2016sne}; (10) \citet{Sim2012sne}; (11) \citet{Marquardt2015sne}; (12) \citet{Leung2018sne}; (13) \citet{Sim2010sne}; (14) \citet{Shen2018sne}; (15) \citet{Pakmor2010sne}; (16) \citet{Pakmor2012sne}; (17) \citet{Kromer2013sne}. \\
SN$_{\rm cc}$: (a) \citet{romano2010};  (b) \citet{WoosleyWeaver1995}; (c) \citet{Nomoto2013Nucleo}; (d) \citet{Heger2002sne}; (e) \citet{Heger2010sne}; (f) \citet{Sukhbold2016sne}. \\
AGB: ($\alpha$) \cite{karakas2010}.}
\end{table*}





\end{document}